\documentclass[prl,aps,twocolumn,groupedaddress,showpacs,floatfix,preprintnumbers,superscriptaddress,longbibliography]{revtex4-1}
\usepackage{amsmath,amssymb,multirow,bm}
\usepackage[colorlinks,linkcolor=blue,anchorcolor=blue,citecolor=blue,urlcolor=blue,filecolor=black]{hyperref}
\usepackage{graphicx}


\newcommand{\beq}{\begin{equation}}
\newcommand{\eeq}{\end{equation}}
\newcommand{\bea}{\begin{eqnarray}}
\newcommand{\eea}{\end{eqnarray}}

\begin{document}
\title{Nuclear fragmentation reactions as a probe of neutron skins in nuclei}
\author{E.A. Teixeira}
\affiliation{Department of Physics and Astronomy, Texas A\&M University-Commerce,  TX 75429-3011, USA}
\affiliation{Departamento de F\'isica, Instituto Tecnol\'ogico de Aeron\'autica-CTA, 12228-900, S\~ao Jos\'e dos Campos, Brazil}
\email{soberania@gmail.com}

\author{T. Aumann} 
\email{thomas.aumann@tu-darmstadt.de}
\affiliation{Institut f\"ur Kernphysik,  Technische Universit\"at Darmstadt, 64289 Darmstadt, Germany}
\affiliation{GSI Helmholtzzentrum f\"ur Schwerionenforschung GmbH, D–64291 Darmstadt, Germany}
\affiliation{Helmholtz Research Academy Hesse for FAIR, D–64289 Darmstadt, Germany} 

\author{C.A. Bertulani}
\email{carlos.bertulani@tamuc.edu}
\affiliation{Department of Physics and Astronomy, Texas A\&M University-Commerce,  TX 75429-3011, USA}
\affiliation{Institut f\"ur Kernphysik,  Technische Universit\"at Darmstadt, 64289 Darmstadt, Germany}
\affiliation{Helmholtz Research Academy Hesse for FAIR, D–64289 Darmstadt, Germany} 

\author{B.V. Carlson}
\email{brettvc@gmail.com}
\affiliation{Departamento de F\'isica, Instituto Tecnol\'ogico de Aeron\'autica-CTA, 12228-900, S\~ao Jos\'e dos Campos, Brazil}
\begin{abstract}
We investigate  the contributions of various reaction channels to the interaction, reaction, charge-changing and neutron-changing cross sections. The goal is to investigate the relation between microscopic interactions and the symmetry energy component of the equation of state (EoS) of interest for the structure of neutron stars. We have made a comparison of the neutron skins extracted from diverse experimental techniques with those obtained with Hartree-Fock-Bogoliubov calculations with 23 Skyrme and with  8 density-dependent interactions used in the relativistic mean field method. We have shown that no particular conclusion can be drawn on the best EoS in view of the wide range of uncertainty in the experimental data. We have further investigated the prospects of using neutron-changing reactions to assess the isospin dependence of the neutron-skin in neutron-rich nuclei.
\end{abstract}

\maketitle

\section{Introduction}
A neutron star (NS) is a very dense baryonic system, with about 20 times more neutrons than protons and a central density reaching 5 to 6 times the saturation density of the matter  in nuclei on earth, $\rho_0 \simeq 0.16$ fm$^{-3}$. NS  are  very complex systems where all four fundamental forces have an essential role in defining their structure. One  cannot rule out the possibility that  hyperons  and quarks exist in their dense cores. Several astronomical observables such as NS  masses, radii and tidal deformabilities, together with experiments  with nuclei on earth, have provided an insight on the equation of state (EoS) of symmetric and asymmetric nuclear matter. The EoS is the main ingredient for the determination of the basic NS properties \cite{Baym_2018}.

The EoS for asymmetric nuclear matter is a relation between the energy per nucleon $\cal E$, the nucleon density $\rho$, and the asymmetry parameter $\delta =(\rho_n - \rho_p)/\rho$, where $\rho_n$ ($\rho_p$) is the neutron (proton) density. It can be written as
\beq
{\cal E} (\rho, \delta) = {\cal E} (\rho)+ S(\rho)\delta^2 +{\cal O}(\delta^4),
\eeq
where the first term depends only on the total nucleon density  $\rho =\rho_n + \rho_p$. The second term accounts for the deviation from symmetric nuclear matter and higher order corrections on $\delta$ are included in  ${\cal O}(\delta^4)$. 
The so-called symmetry energy term $S(\rho)$ of the EoS is the most uncertain one, especially at high nuclear matter densities. To gain  further insight on the symmetry energy term, it is common to expand it in a Taylor series around the nuclear matter (NM) saturation density, so that
\beq
S (\rho) = J +{L\over 3} \alpha+ {1\over 18} K_{sym} \alpha^2 + {\cal O}(\alpha^3),
\eeq
where $ \alpha = (\rho - \rho_0) / \rho_0$ is the expansion parameter, $J= S (\rho_0)$ is the symmetry energy at saturation density $\rho_0$, $L$ is the slope parameter {  and $K_{sym}$ is the curvature parameter. Here will will only discuss the roles of $J$ and $L$. The relevance of the $K_{sym}$ term has been addressed extensively in the literature (see, e.g., Refs. \cite{HOLT201877,Lattimer-2021,PhysRevC.105.065802}).} Whereas the value of  $J \simeq 30$ MeV is compatible with numerous theoretical predictions, $L=(3\rho \partial S/\partial \rho)_{\rho=\rho_0}$ is still poorly constrained \cite{PhysRevLett.106.252501,PhysRevC.92.064304}. This poses a problem when extrapolating from the knowledge of symmetric $\rho_n \simeq \rho_p$ to asymmetric nuclear matter (ANM) with $\rho_p \simeq 0$. The pressure in homogeneous nuclear matter is extracted from the EoS by using the relation $p(\rho, \delta) = \rho^2d{\cal E}(\rho,\delta)/d\rho$. For asymmetric nuclear matter at saturation density one obtains {  to leading order} an additional pressure from the symmetry energy term, given by $p =L\rho_0/3$. Therefore, the slope parameter $L$ needs to be well determined  for a  good description of a neutron star.

Theoretically, several models have been developed to obtain the EoS of nuclear matter, all of them being constrained by comparison with experiment nuclear observables, such as nuclear masses, charge radii, excitation energies of giant monopole and dipole resonances, particles produced in central nuclear collisions at intermediate energies ($\sim$ a few 100 MeV/nucleon).  { Standard microscopic models like the non-relativistic Skyrme-Hartree-Fock (SHF), or the relativistic mean-field (RMF),  have been used quite successfully to describe nuclear properties \cite{BEINER197529,KOHLER1976301,BARTEL198279,dobaczewski1984hartree,STOITSOV20131592,PhysRevC.40.2834,REINHARD1995467,CHABANAT1997710,CHABANAT1998231,PhysRevC.58.220,PhysRevC.60.014316,PhysRevC.68.031304,GORIELY2005425,PhysRevC.85.035201,PRC76-045801,PRC81-034323,PRC74-034323,PLB671-36,PRC63-044303,PRC55-540,PRC71-024312,PRC69-034319,PRC52-054310}. }

Each of these microscopic models with different interactions yields a distinct  energy density functional or EoS for nuclear matter. However, a universal energy density functional  able to describe most nuclear properties and leading to a consistent EoS, has not yet been found.  In other words,  different NM quantities predicted by these models, such as incompressibility and symmetry energy, have a broad range of values as a function of the NM density, even when many other nuclear  properties are well described for a limited set of nuclei. For example, in Ref. \cite{Bertulani.PRC.100.015802} it has been shown that while the parameter $J$ is approximately constant, around 32 MeV, for about 23 popular Skyrme interactions, the slope parameter $L$ can vary from about 30 MeV to 130 MeV. This result has been known for decades and highlights the fact that the 10 or more parameters used to describe Skyrme interactions can lead to very different EoS, albeit being able to describe very well a limited set of nuclear properties \cite{BURGIO2021103879}.

Numerous experiments using complementary techniques have been designed to constrain the symmetry energy term of the EoS. Astronomical observations have also helped to assess this property of the EoS. Here we will discuss one class of laboratory experiments linking the symmetry energy term to the neutron skin in nuclei.  Experiments assessing the size of the neutron skin in nuclei also use numerous techniques, although an accurate result is still lacking. The neutron skin in a nucleus is often defined in terms of  the difference between its neutron and proton root-mean-squared (rms) radius,  $\Delta r_{np} = \left< r_n^2 \right>^{1/2} - \left< r_p^2 \right>^{1/2}$. It has been widely advertised that neutron skins in nuclei are directly correlated with the slope parameter $L$ of NM (see, e.g., Refs. \cite{vinas2014,HOLT201877}).   A recent experiment \cite{PhysRevLett.126.172502} performed at the Jefferson Laboratory, USA, has looked at the analyzing power for polarized electron scattering due to its parity violating (PV) term in the interactions with a $^{208}$Pb target and found an unexpectedly large neutron skin\footnote{The value and error bar reported in Ref. \cite{PhysRevLett.126.172502} is a combination of two separate measurements, denoted as PREX-1 and PREX-2.}
value of $\Delta r_{np} = 0.283 \pm 0.071$ fm. Intriguingly, this result implies a slope parameter $L=106 \pm 37$ MeV, larger than expected from most microscopic calculations and also from most other experimental results. For example, Ref. \cite{ISI:000293447900005} reported a value of $\Delta r_{np} = 0.156^{ + 0.025}_ {- 0.021}$ fm. Recent astronomical observations from heavy pulsar masses, NICER telescope and LIGO/Virgo laboratory seem to clearly point to much smaller values for  $\Delta r_{np}$ and $L$.

Experimental efforts have also been made to deduce the neutron skin from  fragmentation reactions of relativistic nuclei incident on light nuclear targets. Fragmentation reactions can be thought as occurring in two steps: First nucleons are stripped off the projectile and energy is deposited in the primary fragments. After that, the primary fragments decay by evaporation of nucleons, alpha particles, and even by fission, leading to secondary fragments which are actually the ones reaching the detectors. While the first step can be rather accurately determined by theory, the second one involves numerous assumptions about excitation energies, level densities and optical potentials that are not as easily handled by theory. { A way to avoid most of the complications due to the calculation of the evaporation stage is the measurement of total (a) charge- or (b) neutron-changing cross sections, i.e, all the fragments with (a) charge, or (b) isotones, different than the projectile \cite{Blank1992,PhysRevLett.107.032502,10.1093/ptep/ptu134,PhysRevLett.113.132501,PhysRevC.94.064604,AumannPRL119.262501}. The idea is that in both cases the cross sections for secondary fragments is nearly the same as those for primary fragments. Assuming that the main evaporation probability is for decay by emission of neutrons, the detection of all possible (a) isotopes, or (b) isotones, yields the same cross sections as for the total primary (a) isotopes, or (b) isotones,  because all decay channels have been accounted for. The method fails when a primary fragment with neutrons (protons) removed decays by proton, or alpha, emission leading to charge changing and neutron changing cross sections that are not the same for primary and secondary fragments.}  

With the radioactive beams now provided in major nuclear facilities and with the increasing accuracy of better-built detectors, experiments with neutron-rich nuclei are now possible, increasing the constraints on the cross section measurements and their viability to study neutron skins in several nuclei, a clear advantage over fixed target experiments. Total neutron changing cross section measurements with good accuracy are now possible with inverse kinematics, allowing the study of the fragmentation of projectiles along an isotopic chain \cite{AumannPRL119.262501}. 

Any experimental method using a reaction to extract an observable sensitive to EoS is model dependent, since ab initio calculations are not available. The determination of this model uncertainty, i.e., the systematic uncertainties of the reaction theory have to be evaluated and should be included in the error budget of quantities extracted from the measured cross sections. 

The Glauber scattering theory, based on a description of the {  nucleus-nucleus} reaction by {  means of} individual nucleon-nucleon scattering is believed to be a good approximation at high beam energies above few hundred MeV \cite{HufnerPRC12.1888,HRB91,CarlsonPRC46.R30,PhysRevC.51.252,BD:2021}. Since we discuss total reaction probabilities, multiple reactions are not important and we start with the Glauber theory in its simplest form in first order without free parameters. The ingredients are point-neutron and -proton densities (whenever possible derived from experimental data), and the NN interaction is implemented by computing the eikonal wave function for an optical potential using the densities and measured pp and pn cross sections. 

In this work, we make an assessment of uncertainties related to the first-order eikonal approximation and the fact that the interaction is introduced as a free nuclear NN interaction. Deviations due to Coulomb recoil, internal Fermi motion of nucleons, as well as due to Pauli blocking are estimated. Calculations are extended up to the 3rd eikonal order.

Even more important is the assessment of deviations from the Glauber approximation due to interactions beyond individual NN scattering. These are nuclear excitations of collective modes, such as  giant resonances, in the scattering process. For heavier nuclei, these processes yield sizable contributions to the cross sections. The cross sections for nuclear- and electromagnetically-induced excitations of collective states are considered here. Such high-lying continuum excitations contribute (for heavy nuclei) to neutron emission and add thus to the total reaction cross section and to the total-neutron removal cross sections. These calculations have systematic uncertainties but can be used as a basis to estimate for the {  limitations} of the Glauber approach. Alternatively, these cross sections can be measured separately in the experiment, as proposed in  \cite{AumannPRL119.262501}, to avoid this theoretical model dependence. 

\section{Cross sections of primary fragments}  
\subsection{Glauber model for fragment production}
Normalizing the projectile $P$ and target $T$ proton (p) and neutron (n) densities to one, the abrasion-ablation model described in Refs. \cite{Bowman73,HufnerPRC12.1888,CarlsonPRC46.R30,PhysRevC.51.252,BD:2021}, yields the following expression for
the probability that a proton from the projectile {\it survives} the collision with the target {  $(N_T,Z_T)$} for a nucleus-nucleus collision with an impact parameter $b$:
\bea
P_p({\bf b})    &=&
\int d^{2}s\;dz \;\rho_p^{P}\left(  z,\mathbf{s-b}\right)
\left[  1-\sigma_{pp}\int dz'\;\rho_p^{T}\left(z',\mathbf{s}\right)\right]^{Z_T} \nonumber \\
&\times& \left[1  -\sigma_{pn}\int dz'\;\rho_n^{T}\left(z',\mathbf{s}\right)  \right]^{N_T} , 
 \label{abrasionp2}
\eea
where, $\sigma_{pp}$ and $\sigma_{pn}$ are the total (minus Coulomb) proton-proton and proton-neutron scattering cross sections, respectively.
An analogous expression holds for the  neutron survival probability by swapping the role of the neutron and proton in the expression above. Notice that  the factors $1-\sigma \int dz (\cdots)$ are small for any nuclear system, and we can safely use the approximation $1-x \sim \exp(-x)$ in the equations above \cite{CarlsonPRC46.R30,PhysRevC.51.252,BD:2021}.

In the abrasion-ablation model, the cross section for the {\it primary production} of a fragment with $N_F$ neutrons and $Z_F$  protons is
\bea
\sigma{(N_F,Z_F)}
&=&
\left(\begin{array}{c}N_P \\N_F\end{array}\right)\left(\begin{array}{c}Z_P \\Z_F\end{array}\right)
\int d^{2}b \  [P_p({\bf b})]^{Z_{F}}[P_n({\bf b})]^{N_{F}} \nonumber \\
&\times& \left[  1-P_n({\bf b})\right]  ^{N_P-N_{F}}\left[  1-P_p({\bf b})\right]  ^{Z_P-Z_{F}}\, .
  \label{abrasion}
\eea

\subsection{Interaction, neutron- and charge-changing  cross sections}

The {\it interaction cross section} is the sum of all channels for which at least one nucleon is removed, that is
\begin{equation}
\sigma_I= \left[ \sum_{N_F=0}^{N_P}\sum_{Z_F=0}^{Z_P}\sigma{(N_F,Z_F)}\right]-\sigma{(N_P,Z_P)},
\end{equation}
where the second term removes the undisturbed projectile from the sum. 
Using the binomial sum, one obtains
\begin{equation}
\sigma_I= \int d^{2}b \  \left[ 1- [P_p({\bf b})]^{Z_{P}}[P_n({\bf b})]^{N_{P}}\right] \label{sigint}.
\end{equation}
Thus, the interaction cross section is the integral of one minus the probability that all protons and neutrons simultaneously survive the collision.

The {\it neutron-changing cross section}  $\sigma_{\Delta N}$ is the cross section to produce all fragments with the same charge as the projectile by removing at least one of its neutrons. It is obtained by replacing $Z_{F}=Z_{P}$ in Eq. (\ref{abrasion}) and summing from $N_{F}=0$ up to $N_{P}-1$. The sum over the binomial coefficients yields 
\bea
\sigma_{\Delta N}&=& \sigma_{\rm all \; n\;  decay\; channels}^{{Z_P \ {\rm survives}}} =   \sum_{N_F=0}^{N_P-1} \sigma(N_F,Z_P)\nonumber \\
& = &
\int d^2 b \left[P_p(b)\right]^{Z_{P}}\left\{[1- \left[P_n(b)\right]^{N_{P}}\right\} . \label{sigmaDN}
\eea
This means that the probability that $Z_P$ protons survive while all possible numbers of neutrons are removed is equal to the probability that all protons survive (irrespective to what happens to any neutron)  minus the  probability that all protons and neutrons survive, simultaneously (i.e, that the projectile remains intact). 
The charge-changing cross section $\sigma_{\Delta Z}$ is obtained by adding all fragments in which at least one proton is removed. This means that  $\sigma_{\Delta Z} = \sigma_I - \sigma_{\Delta N}$.

\subsection{The optical potential, eikonal phase and S-matrices}
The eikonal S-matrices for a projectile proton scattering off a target nucleus is given by
$
S_p({\bf b}) = \exp\left[i\chi_p({\bf b})\right],\label{eik1}
$
where  \cite{BD:2021}
$\chi_p({\bf b}) = -(\hbar v)^{-1}\int dz\; U_p({\bf r})$
is the eikonal phase, $v$ is the projectile velocity, assumed undisturbed in high energy collisions, and $U_p$ is the optical potential for proton scattering. 

For nucleon removal reactions, only the imaginary part of the optical potential is of relevance and it can be related to the nucleon-nucleon cross section and nuclear densities as \cite{BD:2021} 
\bea
U_p({\bf r}) &=& -i{\hbar v \over 2} 
 \int d^{2}s \;\rho_p^{P}\left(  z,\mathbf{s-b}\right)\nonumber \\
 &\times&  \int dz' \left[Z_T \sigma_{pp} \;\rho_p^{T}\left(z',\mathbf{s}\right) 
+N_T \sigma_{np} \;\rho_n^{T}\left(z',\mathbf{s}\right) \right]. \ \ \ 
\label{eik3}
\eea
The eikonal phase-shift is therefore
\bea
\chi_p({\bf b}) &=&  i{1\over 2} 
 \int d^{2}s dz\;\rho_p^{P}\left(  z,\mathbf{s-b}\right) \nonumber \\
 &\times& \int dz' \left[Z_T \sigma_{pp} \;\rho_p^{T}\left(z',\mathbf{s}\right) 
+N_T \sigma_{np} \;\rho_n^{T}\left(z',\mathbf{s}\right) \right], \ \ \ \ \ \ 
\label{eik30}
\eea
and the S-matrix is
\bea
S_p({\bf b}) &=& \exp\left\{ -{1\over 2} 
 \int d^{2}s dz\;\rho_p^{P}\left(  z,\mathbf{s-b}\right) \right. \nonumber \\
&\times& \left. \int dz' \left[Z_T \sigma_{pp} \;\rho_p^{T}\left(z',\mathbf{s}\right) 
+N_T \sigma_{np} \;\rho_n^{T}\left(z',\mathbf{s}\right) \right] \right\}. \ \ \ \ \ \ 
\label{eik4}
\eea
The probability in Eq. (\ref{abrasionp2}) is given by 
$
P_p ({\bf b}) = \left| S_p({\bf b}) \right|^2 $. Analogous expressions are
obtained for neutron removal with the roles of neutrons and protons inverted. 
{  Notice that Eq. (\ref{eik4}) can be deduced on probability grounds without using the optical potential, Eq. (\ref{eik3}). But later we will need the concept of an optical potential to study the higher-order corrections in the eikonal wavefunctions.}

The optical limit (OL) of the Glauber cross sections is obtained when the second term inside brackets in Eq. (\ref{abrasionp2}) is very small so that  $1-x \sim \exp(-x)$.  Then it is straightforward to show that $\sigma_I \rightarrow \sigma_{OL}$ where
\bea
\sigma_{OL} &=& \int d^2b \Bigg[ 1 - \exp\Bigg\{ - Z_P
 \int d^{2}s dz\;\rho_p^{P}\left(  z,\mathbf{s-b}\right) 
 \nonumber \\
 &\times& \int dz' \left[Z_T \sigma_{pp} \;\rho_p^{T}\left(z',\mathbf{s}\right) 
+N_T \sigma_{np} \;\rho_n^{T}\left(z',\mathbf{s}\right) \right] \nonumber \\ 
&-& N_P
 \int d^{2}s dz\;\rho_n^{P}\left(  z,\mathbf{s-b}\right)  \nonumber \\
&\times&  \int dz' \left[Z_T \sigma_{np} \;\rho_p^{T}\left(z',\mathbf{s}\right)
+N_T \sigma_{pp} \;\rho_n^{T}\left(z',\mathbf{s}\right) \right]\Bigg\}\Bigg].  \ \ \ \ \ \
\label{eik8}
\eea

The equations developed above have been used extensively in the literature to describe interaction, reaction, charge- and neutron-changing cross sections. But, in order to achieve the accuracy needed to extract values of neutron skins and constraints on the EoS of nuclear matter, corrections to these expressions must be considered. In the next sections we study most of the relevant corrections known and investigate the sensitivity of the cross sections on each correction. 
 
\section{Corrections of the fragmentation cross sections}

\subsection{Medium corrections of NN cross sections}
Here we discuss the nucleon-nucleon cross sections used as input in Eqs. (\ref{abrasionp2}-\ref{eik8}). To account for medium effects, we use the Pauli-blocking corrections for nucleus-nucleus collisions extensively discussed in Refs. \cite{Bertulani1884,HRB91,BertulaniJPG01,BertulaniConti10}. 
The reduction of the nucleon-nucleon cross section in the medium due to Pauli-blocking is considered as given by
\begin{eqnarray}
\sigma_{NN}^{Pauli}(E,\rho_1,\rho_2) &=&\sigma_{NN}^{free}(E){1 \over 1+1.892\left({\displaystyle{|\rho_1-\rho_2|\over \tilde{\rho}\rho_0}}\right)^{2.75}}\nonumber \\
&\times& 
\left\{
\begin{array}
[c]{c}%
\displaystyle{1-{37.02 \tilde{\rho}^{2/3}\over E}}, \ \ \   {\rm if} \ \ E>46.27 \tilde{\rho}^{2/3},\\ \, \\
\displaystyle{{E\over 231.38\tilde{\rho}^{2/3}}},\ \ \ \ \  {\rm if} \ \ E\le 46.27 \tilde{\rho}^{2/3},\end{array}
\right.
\label{VM1}
\end{eqnarray}
where $E$ is the laboratory energy in MeV, $\tilde{\rho}=(\rho_1+\rho_2)/\rho_0$,  with $\rho_0=0.16$ fm$^{-3}$, and $\rho_i(r)$ is the local density at position $r$ within nucleus $i$.  The free nucleon-nucleon cross sections were adopted from Ref.  \cite{BertulaniConti10} where a parametrization was developed to fit the experimental data as a function of the energy.

For proton-nucleus collisions, the Pauli blocking correction reads \cite{Clementel1958}
\begin{eqnarray}
\sigma_{NN}^{Pauli}(E,\rho_2) =\sigma_{NN}^{free}(E)\left( 1-{8\over 5}{E_F \over E}\right),   {\rm for}\  E\geq {8\over 5} E_F, \nonumber \\
\label{VM2}
\end{eqnarray}
where $E_F = \hbar^2[3\pi^2\rho_2(r)/2]^{2/3}/2m_N$, $\rho_2$ is the total target nucleon density, with $\rho_1({\bf r}) = \delta({\bf r})$.  Equation (\ref{VM2})  is derived under the assumption that the nucleon-nucleon cross section has an $1/E$ dependence which is a good approximation for $E<100$ MeV. As we show later, for larger energies where $\sigma_{NN}^{free} \sim const.$, the Pauli-blocking contribution to proton-nucleus collisions is small and the exact form of Eq. (\ref{VM2}) becomes irrelevant. 

The average NN cross section at the distance of closest approach between the projectile and the target is obtained using
\begin{equation}
\left\langle \sigma_{NN} (E,b) \right\rangle_{Pauli} = \frac{\int d^3 r  \rho_1({\bf r}) \rho_2({\bf r}+{\bf b}) \ \sigma_{NN}^{Pauli} (E, \rho_1,\rho_2)}{\int d^3 r \rho_1({\bf r}) \rho_2({\bf r}+{\bf b})} , \label{avesigEb}
\end{equation}
where ${\bf b}$ is the impact parameter vector, perpendicular to the beam axis.

To assess the relevance of Pauli-blocking in collisions at low energies,  we consider first the p + $^{208}$Pb interaction cross section. For $^{208}$Pb  we generate the density using a Skyrme SLy5 interaction in the Hartree-Fock-Bogoliubov (HFB) approximation  \cite{RS04}.   The Skyrme force is a function of numerous terms with contact interactions  accounting for coordinate, spin, and isospin dependence. Numerical methods have been developed to calculate nuclear binding energies and other nuclear properties such as the energy density functional $E[\rho]$. To each Skyrme interaction, we added a mixed pairing interaction of the form
\beq
v({\bf r}, {\bf r}') = v_0 \left( 1-{\rho \over 2\rho_0}\right)\delta({\bf r} -{\bf r}') \label{pairing}
\eeq
where $\rho(r) = \rho_n(r) + \rho_p(r)$ is the isoscalar local density. The pairing strength adopted is the same for neutrons and protons, $v_0 = -131.6$ MeV, and the saturation density is fixed at $\rho_0 = 0.16$ fm$^{-3}$.

\begin{figure}[t]
\begin{center}
\includegraphics[
width=3.in
]
{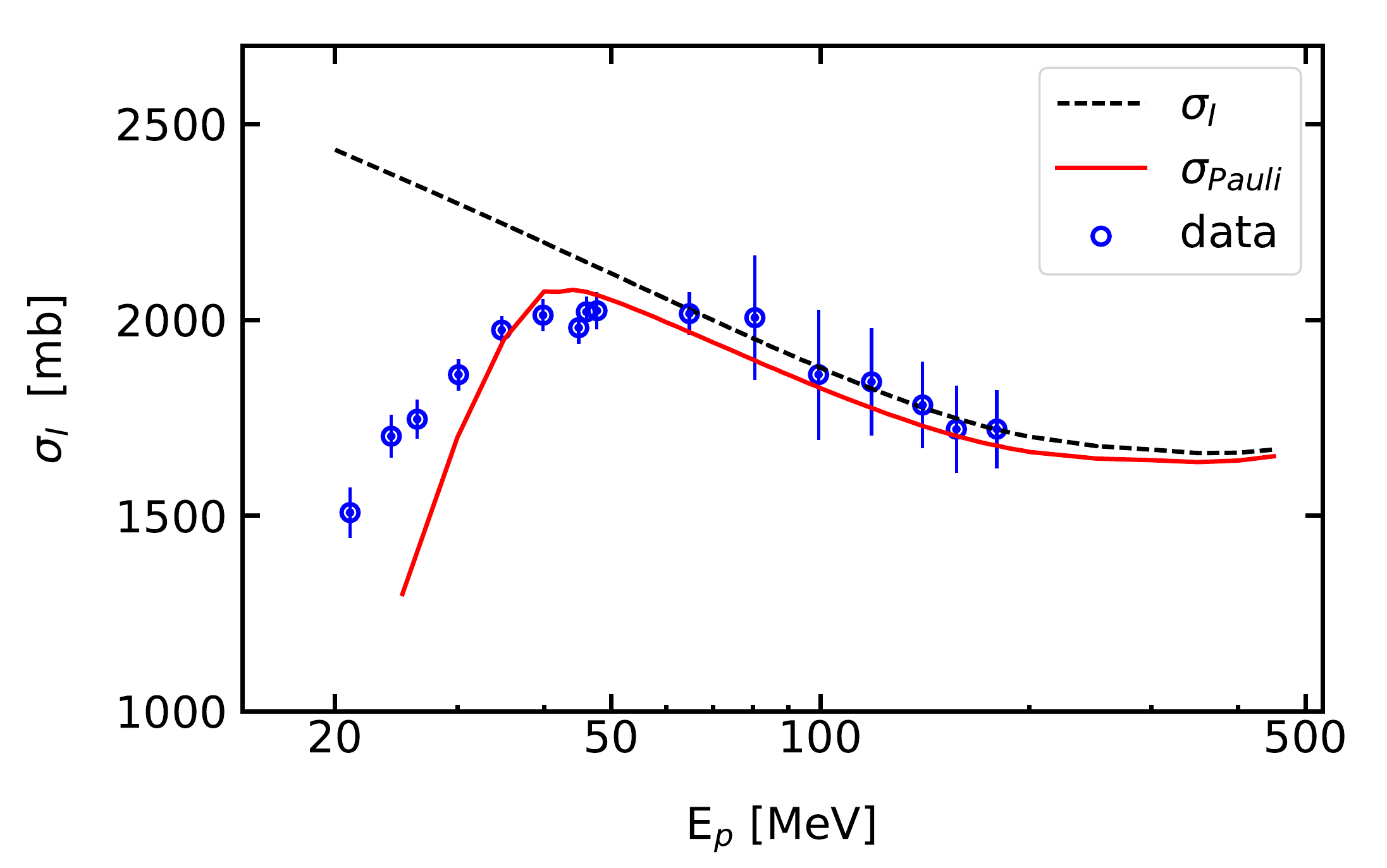}
\caption{Interaction cross sections for p + $^{208}$Pb scattering. Data are from Refs. \cite{PhysRevC.12.1167,INGEMARSSON1999341,PhysRevC.71.064606}. The solid curve includes the effect of Pauli blocking while the dashed curve does not.}
\label{Pauli1}
\end{center}
\end{figure}

The energy dependence of the cross section is shown  in Figure \ref{Pauli1} with data collected from Refs. \cite{PhysRevC.12.1167,INGEMARSSON1999341,PhysRevC.71.064606}.  The solid curve, denoted $\sigma_{Pauli}$, includes the effect of Pauli blocking, while the dashed curve, denoted $\sigma_I$, does not. The effect of Pauli blocking in  $\sigma_{Pauli}$ is to reduce the nucleon-nucleon cross sections in the medium and consequently the probability for nucleon removal. The interaction cross section is therefore also reduced. It is noticeable from Figure \ref{Pauli1} that the inclusion of Pauli-blocking becomes important at  proton bombarding energies below 50 MeV, leading to a better agreement with the experimental data.  It is also visible that the effect is much reduced at large proton bombarding energies beyond a few hundreds MeV. At the very low energies, below $E_p \sim 30$ MeV the calculations taking into account the Pauli blocking effect run into problems, because the equations erroneously  imply that the nucleus becomes fully transparent at the large central densities. This is an overestimation of the effect of Pauli blocking. In this limit, an appropriate Brueckner g-matrix method (see, e.g., Ref. \cite{PhysRevC.104.024606}) would be more appropriate than the approach adopted here. However, the simpler method adopted here yields similar results to those using the g-matrix method reported in Ref. \cite{PhysRevC.104.024606}.

\begin{figure}[t]
\begin{center}
\includegraphics[
width=3.in
]
{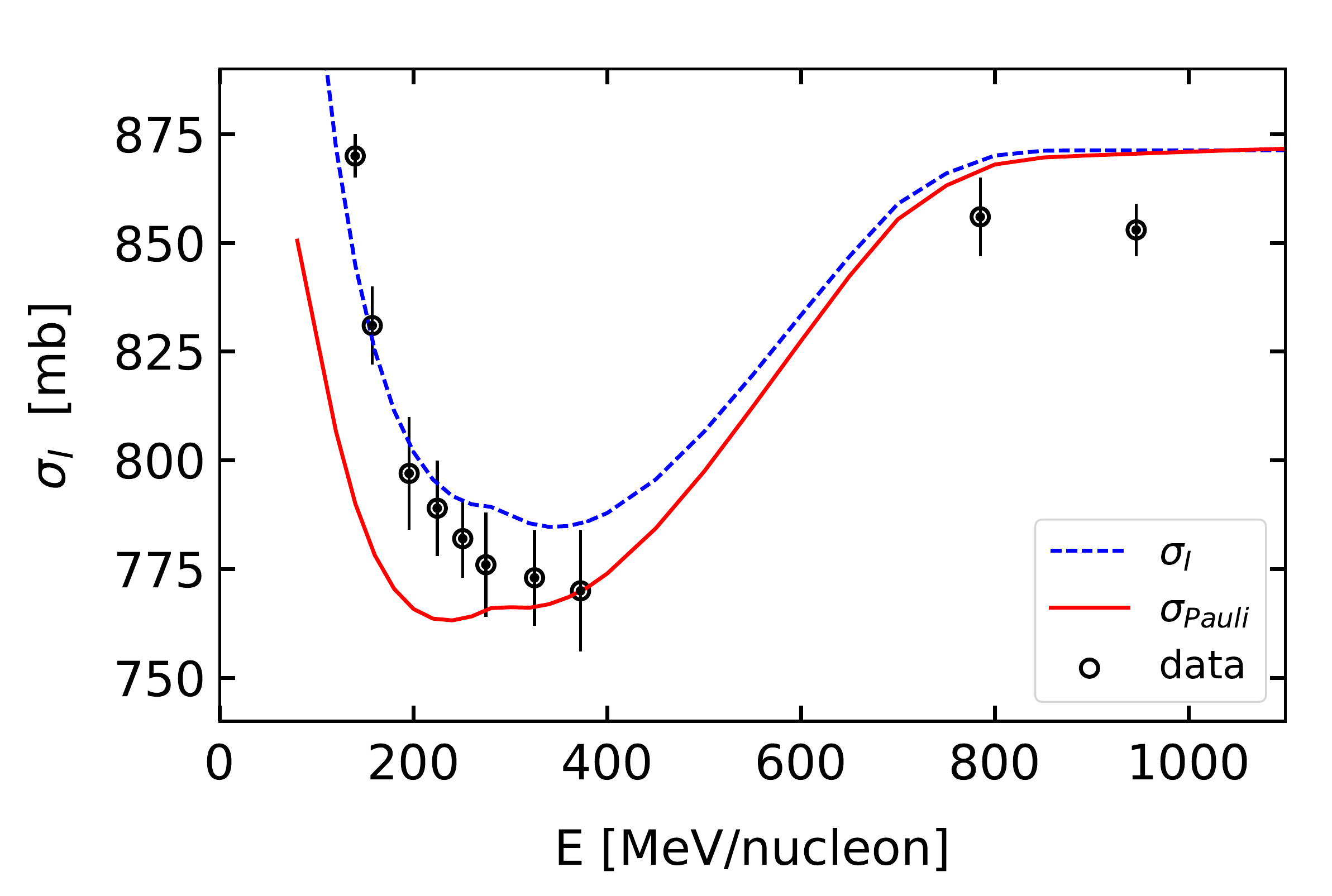}
\caption{Interaction cross section for $^{12}$C + $^{12}$C collisions as a function of the bombarding energy in MeV/nucleon. The solid (dashed) curve describes the cross sections (Eq. (\ref{sigint})) calculated with (without) Pauli blocking corrections.}
\label{Pauli}
\end{center}
\end{figure}

For nucleus-nucleus collisions, we consider  the interaction cross section for $^{12}$C + $^{12}$C reactions as a function of the bombarding energy in MeV/nucleon. This is shown in Figure \ref{Pauli} with data collected from Ref.  \cite{PhysRevC.79.061601,OZAWA200132}. The calculations were made with charge densities extracted from electron scattering, as also used in  Ref. \cite{PhysRevC.79.061601}. The same profile density was assumed for neutrons and for protons. Calculations were performed with Eq. (\ref{sigint}) which yields the same results as the optical limit cross sections, given by Eq. (\ref{eik8}). 

Intriguingly, and contrary to expectations, inclusion of Pauli blocking does not appear improve the comparison with the experimental data at low bombarding energies, as seen in Figure \ref{Pauli}. The data presented in this figure also seem to be in apparent contradiction with the proton-nucleus collision data presented in Figure \ref{Pauli1}, because the inclusion of Pauli-blocking  at low energies seems to play an important role in the proton-nucleus case.  The experimental data presented in Figure \ref{Pauli} are also at odds with the studies reported in Refs. \cite{HRB91} for $^{12}$C + $^{12}$C collisions. We have not found a good explanation for this inconsistency. Notice that the lowest data point in Figure  \ref{Pauli} is at $E = 33 $ MeV/nucleon. Looking at Figure \ref{Pauli1} we see that this is not much below the point of inflection of the solid curve which includes the medium effect.  Thus it might seem natural to think that Pauli blocking effects should not play an important role at the energies of the data of Figure \ref{Pauli}. However, as reported in Refs.  \cite{Bertulani1884,HRB91,BertulaniJPG01,BertulaniConti10}, the inclusion of two occupied Fermi spheres in momentum space also leads to an expected larger reduction of nucleon-nucleon cross sections in nucleus-nucleus collisions at this energy.

\subsection{Fermi motion}

{  Now we discuss the effect of Fermi motion on the nucleon-nucleon cross sections. This effect has been studied previously using different methods than the one we describe here. A few examples are given in Refs. \cite{GOLDHABER1974306,Bertocchi1974,LynchARNPS87,Fan2016FermimotionEO,PhysRevC.79.061601}.} For a given impact parameter, the Fermi motion of nucleons inside the projectile and target modify the collision momentum between the nucleons, leading to an effective momentum $p_{eff} = p_0 + \Delta p$, with the unmodified collision energy $E=p_0^2/2m_N$, and $\Delta p$ denoting the additional momentum due to Fermi motion at the point of collision in the overlap region between the nuclei. We define the Fermi motion averaged NN cross section  as
\begin{equation}
\left\langle \sigma_{NN} (b) \right\rangle_{Fermi} = \frac{\int d^3 r  \rho_1({\bf r}) \rho_2({\bf r}+{\bf b}) \sigma_{NN}^{Fermi} (p_0, {\bf r})}{\int d^3 r \rho_1({\bf r}) \rho_2({\bf r}+{\bf b})} , \label{avesigEb2}
\end{equation}
where the local momentum averaged cross section is 
\begin{equation}
\sigma_{NN}^{Fermi} (p_0, {\bf r}) ={1\over 2\Delta p} \int_{p_0-\Delta p}^{p_0+\Delta p} dp \, \sigma_{NN}^{free} (p) , \label{avesigEb2}
\end{equation}
and $\Delta p = p_{F_1}+p_{F_2}$, with the local Fermi momenta $p_{F_i} = \hbar [3\pi^2\rho_i(r)/2]^{1/3}$, where $\rho_i(r)$ are the total projectile ($i=1)$ and target ($i=2$) densities at position ${\bf r}$. The Fermi motion averaged interaction cross section $\left\langle \sigma_{NN} (b) \right\rangle_{Fermi}$ is then provided as input in Eqs. (\ref{abrasionp2}-\ref{eik8}). 

We have found that the Fermi motion increases the interaction cross section for $^{12}$C + $^{12}$C collisions, albeit the change is very small. At 30 MeV/nucleon the Fermi motion modified cross section, $\sigma_{FM}$, is about 3\% larger than the interaction cross section without inclusion of Fermi motion. However, this decreases dramatically as the energy increases, becoming less that 0.1\% at energies larger than 200 MeV/nucleon.  This is a physically reasonable result, as the net effect of Fermi motion is to increase the interaction cross section at low energies. The reason for the slight increase at low energies is that the NN cross section decreases as $1/E$, thus favoring the lower relative energies created by the Fermi motion.  The  same trend was observed in Ref.  \cite{PhysRevC.79.061601} where Fermi motion was included using an average based on the Goldhaber model for the internal momentum distribution. We find a smaller correction due to the Fermi motion than reported by those authors. The reason for this difference probably lies in the different prescriptions used to tackle the Fermi motion problem. Notice that our method inherently washes out the effect of the Fermi motion, due to nucleons colliding symmetrically around the bombarding momentum $p_0$ along the beam axis, as clearly displayed in Eq. (\ref{avesigEb2}). It is also clear that the slight increase of the cross sections due to Fermi motion is not sufficient to offset the decrease induced by the Pauli blocking effect. 

\subsection{Higher-order eikonal corrections}

The first-order eikonal approximation neglects higher derivatives of the optical potential. In Refs. \cite{PhysRevLett.27.622,WALLACE1973190}, Wallace has investigated analytical corrections arising from higher-order derivatives in the eikonal S-matrices. To increasing order it was found that the total S-matrix can be cast as a sum of the form
\beq
S(b) = S^{(\mathrm{0})}(b) +  S^{(\mathrm{LO})}(b) + S^{(\mathrm{NLO})}(b) + S^{(\mathrm{NNLO})}(b) + \dots,
\eeq
with  the solutions
\bea
S^{(\mathrm{0})}(b)&=& \exp[i\chi_0 (b)] \nonumber \\
S^{(\mathrm{LO})}(b)&=& S^{(\mathrm{0})}(b)\exp[i\chi_1 (b)] \nonumber \\
S^{(\mathrm{NLO})}(b)&=& S^{(\mathrm{LO})}(b)\exp \left[i\chi_2(b)-\psi_2 (b)\right] \nonumber \\
S^{(\mathrm{NNLO})}(b)&=& S^{(\mathrm{NLO})}(b)\exp \left\{i\left[\chi_3(b)+\lambda_3(b)\right]+\psi_3(b)\right\}, \nonumber \\ \label{scorr}
\eea
where $S^{(\mathrm{0})}(b)$ is the usual eikonal S-matrix, as in Eq. (\ref{eik4}), and we use a notation in which the corrections are to leading-order (LO), next-to-leading order (NLO) and next-to-next-to-leading-order (NNLO). The analytical expressions for the phase shifts entering the modified S-matrices are \cite{PhysRevLett.27.622,WALLACE1973190} 
\bea
\chi_0 (b)&=& -2k\epsilon\int_0^\infty dz \,u(r)  \nonumber \\ 
\chi_1 (b)&=& -k\epsilon^2  (1+\beta_1)\int^\infty_{0} dz\,u^2(r)  \nonumber \\
\chi_2(b)&=&-k\epsilon^3  \left(1+\frac{5}{3}\beta_1 +\frac{1}{3}\beta_2\right)\int^\infty_{0} dz\,u^3(r)\nonumber  \\
&-& \frac{b}{24k^2} \left[\chi_0' (b)\right]^3 ,
\label{scorr2}
\eea
and
\bea 
\psi_2(b) &=& \frac{b}{8k^2} \chi_0'(b) \nabla^2\chi_0(b) \nonumber \\
\chi_3(b) &=&-k\epsilon^4 \, \left(\frac{5}{4} +\frac{11}{4}\beta_1 +\beta_2 +\frac{1}{12}\beta_3\right) \int^\infty_{0}  dz \, u^4(r)  \nonumber \\
&-& \frac{b}{8k^2}  \chi_1'(b) \left[\chi_0'(b)\right]^2 \nonumber \\
\lambda_3(b) &=&  -k\epsilon^2 \, \left( 1+\frac{5}{3} \beta_1+\frac{1}{3} \beta_2\right) \int^\infty_{0} dz \left[ \frac{1}{2k} {\partial u(r) \over \partial r}\right]^2  \nonumber \\
\psi_3(b) &=&\frac{1}{8k^2} \left[b \chi_0'(b) \nabla^2 \chi_1(b) +b \chi_1'(b) \nabla^2\chi_0(b) \right].  \label{eikcorr} 
\eea
where $\beta_n = b^n \partial{^n}/{\partial b^n}$, is a dimensionless derivative of order $n$ in the transverse direction,  $u(r)=U(r)/U(0)$, where $U$ is the optical potential, and $\epsilon =U(0)/2E$ can be identified as an expansion parameter. For large bombarding energies, $E$, the higher order corrections should become irrelevant. Typical depths of the optical potential are of the order of 100 MeV, leading to a prediction that these corrections should be small for energies above 100 MeV for proton-nucleus collisions and at smaller energies for nucleus-nucleus collisions. 

The corrections of the eikonal S-matrix as proposed by Wallace have been incorporated in recent studies of one-nucleon knockout reactions in Ref. \cite{PhysRevC.90.024606} and for elastic scattering and breakup reactions involving halo nuclei in Ref.  \cite{PhysRevC.96.054607}. In both cases, it has been shown that higher-order eikonal corrections are only relevant below 50 MeV/nucleon. Here we investigate if these corrections could modify the fragmentation cross sections appreciably. Higher-order eikonal corrections  should apply to protons and neutron removal probabilities, adding to Eqs. (\ref{eik30}) and (\ref{eik4}). Due to the derivatives connected to $\beta_n$ the corrections are expected to be larger at the nuclear surface. The usual eikonal approximation is thus valid if $ka\gg 1$, where $a$ is the nuclear diffuseness. The higher order corrections may become important for the nuclear potential, but are irrelevant and cancel out for the Coulomb potential which decreases with $1/r$~\cite{PhysRevLett.27.622,WALLACE1973190}.

A few remarks are in order before we proceed. Since the ``fragmentation optical potential'' represented by Eq. (\ref{eik3}), and the corresponding one for its neutron $U_n$ counterpart, are purely imaginary,  $u(r)$ in the equations above is real and  $\chi_0$ and $\epsilon$ are  imaginary. Moreover, since the nucleon removal probabilities are obtained from the product $SS^*$, the additional phases introduced in  $S^{(\mathrm{LO})}$ or $S^{(\mathrm{NNLO})}$ do not change the removal probabilities. Additionally, $\psi_2(b)$ does not contribute to the  fragment production cross sections. These claims can be checked by inspection of Eqs. (\ref{scorr}) and (\ref{eikcorr}). In summary, for the purposes of calculating the isotope production and interaction cross sections, the only higher-order correction needed for the eikonal phase is the addition of the phase  $\chi_2$ in equation (\ref{scorr2}) to the standard eikonal phase $\chi_0$.  Thus, it is naturally expected that higher-order eikonal corrections will be smaller in the case studied here than those reported in Refs. \cite{PhysRevC.90.024606,PhysRevC.96.054607}

In Table \ref{tab1} we show in the second and third columns the interaction cross section for $^{12}$C + $^{12}$C collisions at selected bombarding energies. In the third column we show the interaction cross section including higher-order eikonal corrections, $\sigma_{eik}$. These corrections are small, except for energies below 100 MeV/nucleon, and even in this case they do not comprise more than a 1\% correction.

\begin{table}[ht]
\begin{center}
\caption{Interaction cross sections $\sigma_I$ for $^{12}$C + $^{12}$C collisions at selected bombarding energies.  $\sigma_{heik}$ denotes interaction cross sections including higher-order eikonal corrections, while $\sigma_{Coul}$ denotes interaction cross sections including Coulomb repulsion. \label{tab1}}
\begin{tabular}{|c|c|c|c|}
\hline\hline
E (MeV/nucl) & $\sigma_I$ & $\sigma_{heik}$   &  $\sigma_{Coul}$    \\ 
\hline 
50 & 1139 & 1133 & 1026 \\
\hline 
70 & 980 & 977 & 969 \\
\hline
100 & 909.3 & 908 & 902  \\
\hline
200 & 801.9 &  802.0 & 796         \\
\hline
500 & 806.6 &  806.6 & 804         \\
\hline
\hline
\end{tabular}
\end{center}

\end{table}

\subsection{Coulomb repulsion correction}

At low collision energies one must add a correction due to the Coulomb deflection of the nuclear trajectory. This correction amounts to replacing the impact parameter variable within the integral with  $b'$ due to repulsion as nuclei pass by at closest distance. That is, \cite{BD:2021}
\beq
 b'=a_0+ \sqrt{a_0^2+b^2},
 \eeq 
 where $a_0 = Z_PZ_T e^2/(\gamma \mu v^2)$ is half the distance of closest approach in a head-on collision of point charged particles. {  $\gamma = (1-v^2/c^2)^{-1/2}$ is the Lorentz contraction factor and $\mu$ is the reduced mass of the system.}
This correction leads to an improvement of the eikonal amplitudes for the scattering of heavy systems in collisions at low energies (see Ref. \cite{BD:2021} for more details).

The fourth column in Table \ref{tab1}  shows the cross section corrected by Coulomb recoil. Both the higher-order eikonal correction and the Coulomb repulsion correction are small. But the Coulomb recoil affects the cross section more, specially at lower energies.  The Coulomb recoil increases the closest distance between the projectile and the target and the corresponding nucleon removal probability also decreases.

\subsection{Relativistic corrections}
{  It is evident that relativistic corrections are very important in nuclear reactions at high energies as low as 100 MeV/nucleon where the nucleon mass has already increased by about 10\%. This is very important for inelastic processes, such as the excitation of giant resonances, discussed in the next section. The corrections are not only related to kinematics, but also in the reaction dynamics. Except for a few studies \cite{BertulaniPRL2005}, this important issue has been largely ignored in the literature. Notably, the basic concept of optical potential is not a relativistic one, as one needs a four-potential to comply with proper Lorentz transformations. Besides, simultaneity and retardation are a major theory hurdle when dealing with relativistic many-body systems \cite{BertulaniPRL2005,PhysRevC.83.024907}. Therefore, the concept of an optical potential, as displayed by Eq. (\ref{eik3}) lacks the proper relativistic treatment. In this work we use the optical potential concept to relate it to the calculation of the eikonal phase, Eq. (\ref{eik30}), and the eikonal S-matrix, Eq. (\ref{eik4}), although they can also be deduced on probabilistic grounds.  The optical potential of Eq. (\ref{eik3}) is also a necessary condition to calculate the higher order eikonal corrections based on the adopted formalism from Refs.  \cite{PhysRevLett.27.622,WALLACE1973190}. 

Glauber models depend on the transverse coordinates of the colliding systems, usually identified as impact parameters. These coordinates are Lorentz covariant, the probabilities and cross sections being thus unchanged by Lorentz transformations. However, in Eq. (\ref{abrasionp2}) and the following ones contain integrations along the longitudinal direction. Longitudinal directions are not Lorentz invariant, as they contract along the direction of motion. This amounts to change $dz \rightarrow dz/\gamma$ and $\rho({\bf b},z) \rightarrow \gamma \rho({\bf b},z)$ and the Lorentz factor cancels out. This comes at no surprise because our adopted Glauber procedure amounts to calculate the probability of a binary collision along a tube of thickness $d^2b$ in the direction of motion. The number of nucleons within the tube is unchanged by Lorentz contraction.}

\subsection{Reaction versus interaction cross section}

The excitation of nuclear states leading to decay by nucleon emission can also contribute to the fragmentation cross sections. Adding these contributions to the interaction cross section yields the so-called reaction cross section. The major contribution comes from the excitation of giant resonances (GR), a highly collective excitation mode of the nucleus. Denoting this contribution by $\sigma_{GR}$, the reaction cross section is $\sigma_R = \sigma_I + \sigma_{GR}$. The excitation of giant resonances through the nuclear and Coulomb interaction between the nuclei has been extensively discussed in past publications. In Ref. \cite{Bertulani.PRC.100.015802} it was studied in the same context as the one discussed in this work, namely its contribution to fragmentation reactions through the excitation and decay of giant resonances, mainly by neutron emission. Here we adopt the same method as in Ref. \cite{Bertulani.PRC.100.015802} to calculate the nuclear and Coulomb excitation cross sections {  with relativistic corrections included}.

\begin{figure}[t]
\begin{center}
\includegraphics[
width=3.in
]
{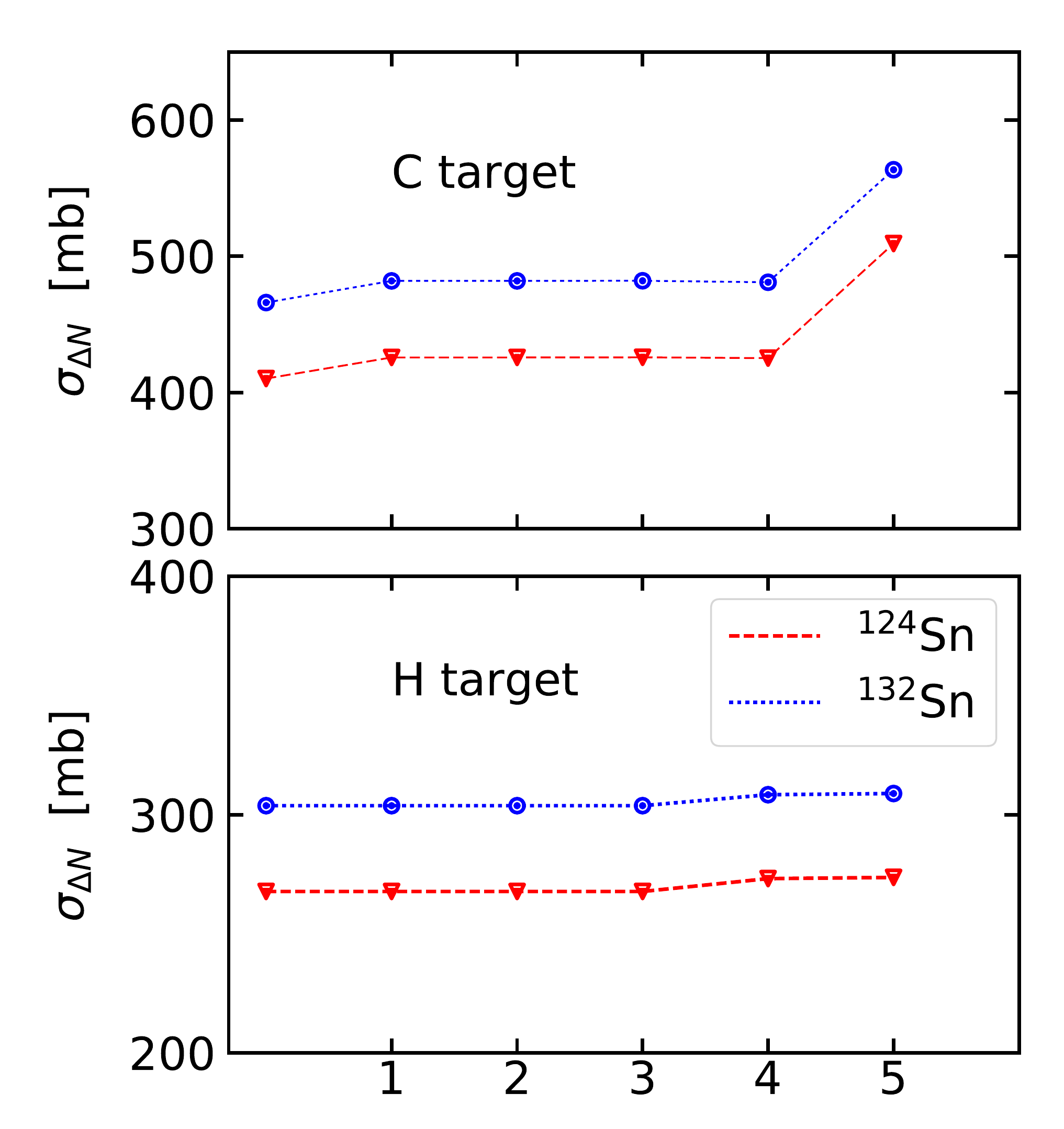}
\caption{Corrections of the neutron removal cross sections for $^{124}$Sn and $^{132}$Sn incident on carbon and hydrogen targets at 900 MeV/nucleon. Corrections include Pauli blocking (1), Coulomb recoil (2), higher order eikonal corrections (3), Fermi motion (4), and Coulomb and nuclear excitation of giant resonances (5). The lines are guides to the eye.}
\label{Corr}
\end{center}
\end{figure}

Since the motivation of this work is to study the effect of the neutron skin along an isotopic chain via the analysis of experiments with neutron rich nuclei at high energies, we consider here the fragmentation cross sections of heavier nuclei incident on proton and carbon targets. In Figure \ref{Corr} we show the contribution of the several corrections of the neutron removal cross sections for $^{124}$Sn and $^{132}$Sn incident on carbon and hydrogen targets at 900 MeV/nucleon. {  Here we use the SLY4 \cite{CHABANAT1998231} interaction to obtain the $^{124}$Sn and $^{132}$Sn proton and neutron densities.} Corrections include Fermi motion (1), Coulomb recoil (2), higher order eikonal corrections (3), Pauli blocking correction (4), and Coulomb  and nuclear excitation of giant resonances (5). The lines are guides to the eye. The parameters used for the excitation of giant resonances are the same as reported in  Ref. \cite{Bertulani.PRC.100.015802} and we assume that they decay by neutron emission only. It is evident that all corrections, except for the last two (Pauli blocking and excitation of giant resonances) are very small at this bombarding energy. For proton targets, the Pauli correction is less than 2\% and the contribution of nuclear and  Coulomb excitation cross section is below 0.3\%.  However, for carbon targets, the nuclear excitation cross sections are of the order of 60-80 mb, making an important contribution to the neutron removal reaction cross sections.

\subsection{Secondary binary collisions}
{  As the mass of the fragment nucleus increases, secondary collisions of abraded nucleons with others in the fragment become an increasingly important source of additional abraded nucleons. This is a simple consequence of the increase in the thickness of the matter that an abraded nucleon must traverse to escape the fragment. These secondary collisions occurring after the abrasion stage are not taken into account in the Glauber formalism adopted here. However, due to the smaller energy of the secondary nucleons and the higher Coulomb barrier of a heavier fragment, the secondary emitted particles are predominantly neutrons. They thus tend to broaden the mass distribution of each charge state of the fragment without changing the charge distribution. The total cross section for each charge state, in particular, the total neutron removal cross section, is therefore relatively immune to this process. We do not consider such secondary binary collisions in this work.} 

\section{Testing neutron densities from mean-field models}
\subsection{Comparison to fragmentation reactions}
 We compare  our calculations of reaction cross sections, $\sigma_R$, charge changing cross sections, $\sigma_{\Delta Z}$, and neutron removal cross sections, $\sigma_{\Delta N}$, to several experiments using nuclear densities generated using the nonrelativistic Hartree-Fock-Bogoliubov (HFB) approximation with  23 different Skyrme interactions and the relativistic mean field (RMF) approximation using 6 nonlinear and 2 density-dependent interactions. The Skyrme interactions used are the  SIII \cite{BEINER197529}, SKA and SKB \cite{KOHLER1976301},  SKM* \cite{BARTEL198279}, SKP \cite{dobaczewski1984hartree}, UNE0 and UNE1 \cite{STOITSOV20131592}, SKMP \cite{PhysRevC.40.2834}, SKI2, SKI3, SKI4 and SKI5 \cite{REINHARD1995467}, SLY230A \cite{CHABANAT1997710}, SLY4, SLY5, SLY6, and SLY7 \cite{CHABANAT1998231},  SKX \cite{PhysRevC.58.220}, SKO \cite{PhysRevC.60.014316},  SK255 and SK272 \cite{PhysRevC.68.031304}, HFB9 \cite{GORIELY2005425} and SKXS20 \cite{PhysRevC.85.035201}. The calculations for the ENE0 and UNE1 interactions performed done with the modified version of the code HFBTO code \cite{STOITSOV20131592}. To all Skyrme interactions a pairing force was added with the form described in Eq. (\ref{pairing}). The RMF calculations used the six nonlinear interactions BSR6 and BSR14 \cite{PRC76-045801,PRC81-034323}, FSUZG00 \cite{PhysRevLett.95.122501,PRC74-034323}, NL3* \cite{PLB671-36}, NLRA1 \cite{PRC63-044303}, NL3-II \cite{PRC55-540}, and the two density-dependent interactions DDME2 \cite{PRC71-024312} and PKDD \cite{PRC69-034319}. The RMF calculations used the effective pairing force described in Ref. \cite{PRC52-054310}. 
 
{  The large number of  interactions covered in this work encompasses a wide range of nuclear matter properties.  A sample of the variation of the parameters of the symmetric matter EOS is shown for  few of these interactions in Table \ref{NMp}. It is worthwhile noticing that the slope parameter $L$ is the most uncertain of all quantities. }

\begin{table}[ht]
\begin{center}
\caption{{  For a few interactions used in this work, we show a sample of the nuclear matter properties at saturation density. All quantities are in MeV units. \label{NMp}} }
\begin{tabular}{|c|c|c|c||c|c|c|c|c|}
\hline\hline
Skyrme & $K_{0}$ & $J$   &  $L$ & Skyrme & $K_{0}$ & $J$   &  $L$   \\ 
\hline 
SIII & 355. & 28.2 & 9.91&  DDME2 & 251. & 32.3 & 51. \\
\hline 
SKP & 201. & 30.0 & 19.7& FSUG00 & 240. & 31.43 & 62.19 \\
\hline 
SKX & 271. & 31.1 & 33.2 & SKXS20 & 202. & 35.5 & 67.1\\
\hline
HFB9 & 231. &  30.0 & 39.9   & SKO & 223. & 31.9 & 79.1 \\
\hline
SLY5 & 230. & 32.0 & 48.2   & SKI5 & 255. & 36.6 & 129.      \\
\hline
\hline
\end{tabular}
\end{center}
\end{table}

\begin{figure}[t]
\begin{center}
\includegraphics[
width=3.in
]
{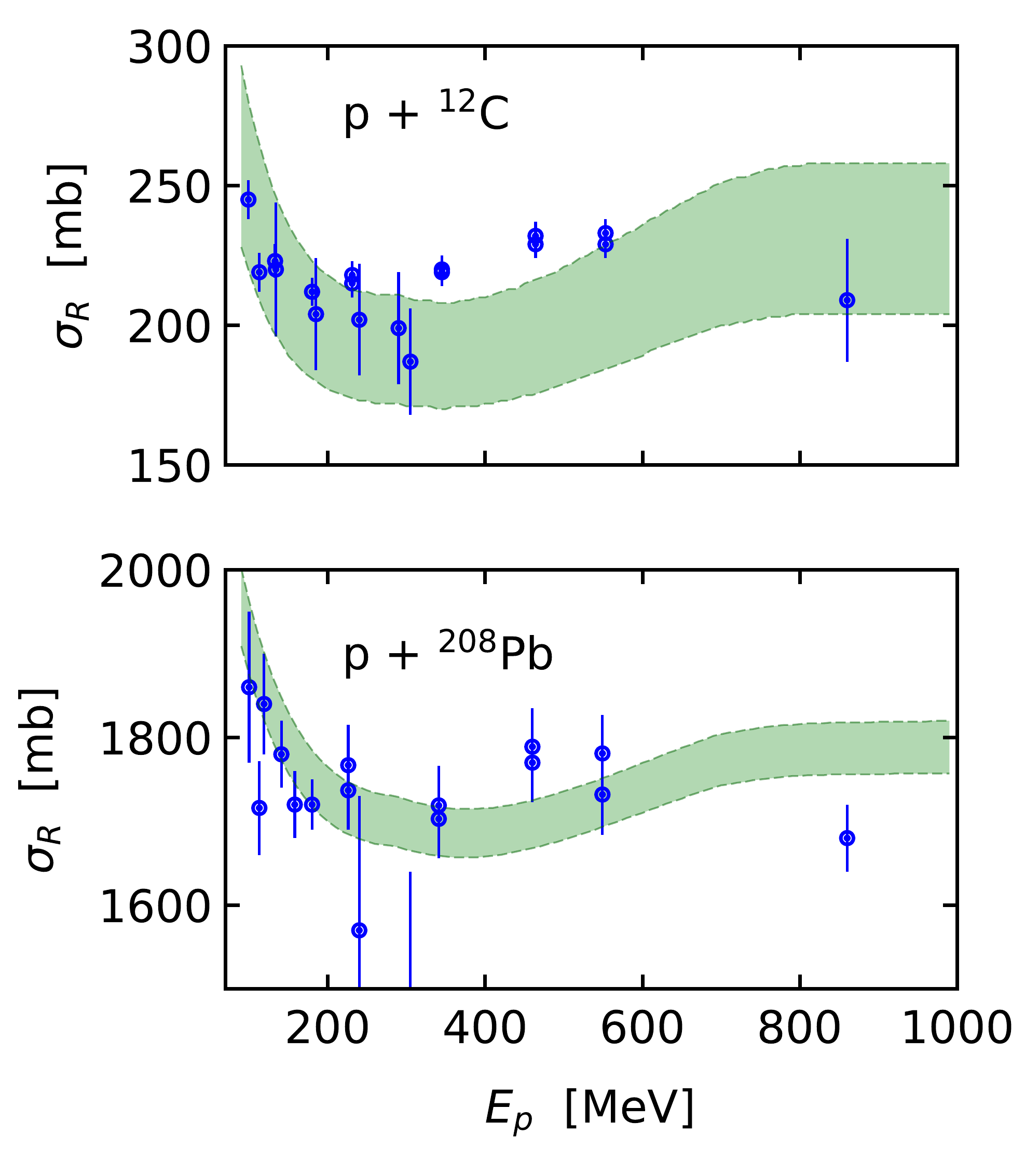}
\caption{Calculated total reaction cross sections for p$+^{12}$C (upper panel) and p$+^{208}$Pb (lower panel) collisions compared to experimental data from Refs. \cite{BAUHOFF1986429,PhysRevC.71.064606}. The shaded band represents the spread of calculations using densities obtained with non-relativistic and relativistic mean-field calculations using a total of 31 different interactions.}
\label{sigmarcpb}
\end{center}
\end{figure}

We consider reactions in the energy range from 100 MeV/nucleon to 1 GeV/nucleon, including corrections due to Pauli blocking, Fermi motion, eikonal higher-order corrections, and Coulomb recoil. These are small corrections, at the level of 2\% or less, and we can easily quantify them with the theory discussed in the previous sections. The most relevant additions to the interaction cross section are due to the excitation of giant resonances, discussed in the last section. In Figure \ref{sigmarcpb} we compare the results of our calculations with total reaction cross sections for p$+^{12}$C (upper panel) and p$+^{208}$Pb (lower panel) collisions compared to experimental data from Refs. \cite{BAUHOFF1986429,PhysRevC.71.064606}. The shaded bands represent calculations using densities obtained with non-relativistic and relativistic mean-field calculations with the 31 different interactions.   {  We arbitrarily select the five best interactions based on the smallest chi-square fit to the data. For the data presented in the upper panel of Fig. \ref{sigmarcpb} they are the DDME2, SKO, SKX, SLY230A and SKXS20, yielding an average slope parameter of $\left<L\right> = (54.9 \pm 23)$ MeV, where the error reflects the range of values obtained from the interactions. The nucleus $^{12}$C is not expected to have a sizable neutron skin and the neutron and proton densities should be well described by the electron scattering data reported in Ref. \cite{PhysRevC.79.061601}. The large variation of the reaction cross sections suggests that the density calculations using mean field methods is not appropriate for such a light nucleus. In fact, the lower panel of Fig. \ref{sigmarcpb} shows that for a large nucleus the calculated mean field densities are more uncertain allowing for a better inference of the neutron skin by a comparison with experimental data. In this case, best $\chi^2$ comparisons with the data are obtained with the BSR14, SLY4, SKA, SKI4, and UNE0. They yield an average slope parameter $\left<L\right> = (59.5 \pm 14.1)$ MeV and an average neutron skin of $\left< \Delta r_{np}\right>=(0.188\pm 0.021)$ fm for the $^{208}$Pb nucleus. This is in rather good agreement with $\Delta r_{np} = (0.156 \pm 0.023)$ fm reported in the experimental analysis of Ref. \cite{ISI:000293447900005} and substantially smaller than the value  $\Delta r_{np} = 0.283 \pm 0.071$ fm from the average of PREX1 and PREX2 experiments implying $\left<L\right> = (106 \pm 37)$ MeV \cite{PhysRevLett.126.172502}. 

It must be pointed out that the extraction of the neutron skin by comparison with the reaction cross section data presented in the lower panel of Fig. \ref{sigmarcpb} is rather deceiving. It is evident from the figure that the accuracy of the data is not good enough to constrain the best interactions reproducing the data. The limited information we can extract from this comparison is that the energy dependence of the cross section data can be rather well described with the densities from microscopic calculations, except for the highest energy points. Besides, the neutron skin extracted from such a comparison is more a convergence of the mean field calculations toward a narrow range of neutron skins than a true constraint placed by the experimental data.} 

\begin{figure}[t]
\begin{center}
\includegraphics[
width=3.2in
]
{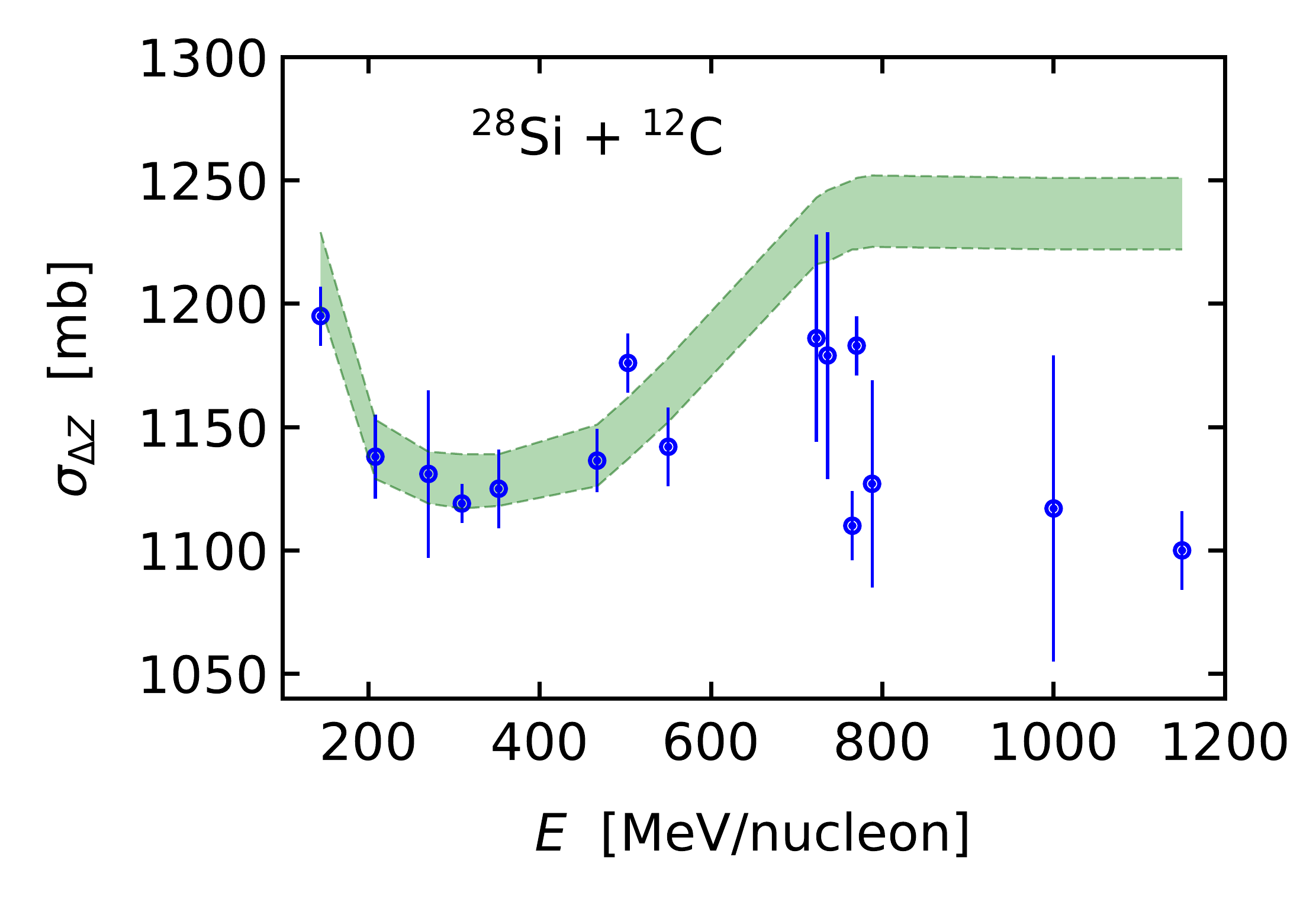}
\caption{Calculated total charge changing cross section, $\sigma_{\Delta Z}$, for the reaction $^{28}$Si + $^{12}$C  compared to experimental data from Refs. \cite{PhysRevC.41.533,FLESCH2001237,ZEITLIN2007341,CECCHINI2008206,PhysRevC.82.014609,LI2016314,Jun-ShengLi:102501}. The shaded band represents densities  {  for $^{28}$Si} obtained with non-relativistic and relativistic mean-field calculations with a total of 31 different interactions.}
\label{Sicc}
\end{center}
\end{figure}

Total charge changing reactions, $\sigma_{\Delta Z}$ \cite{Blank1992,PhysRevLett.107.032502,10.1093/ptep/ptu134,PhysRevLett.113.132501,PhysRevC.94.064604,AumannPRL119.262501}, namely fragmentation reactions in which all isotopes are measured, can be a good probe of neutron skins because, if the measurements are accurate, they can be compared to a simple Glauber calculation of primary fragments, similar to Eq. (\ref{sigmaDN}).  A comparison of our calculations to the experimental data from Refs. \cite{PhysRevC.41.533,FLESCH2001237,ZEITLIN2007341,CECCHINI2008206,PhysRevC.82.014609,LI2016314,Jun-ShengLi:102501} for the reaction $^{28}$Si + $^{12}$C is made in Fig. \ref{Sicc}. The shaded band represents the spread of all calculations done with the densities {  for $^{28}$Si} obtained from non-relativistic and relativistic mean-field calculations using a total of 31 different interactions. {  The $^{12}$C density was taken from electron scattering experiments \cite{PhysRevC.79.061601}}. The comparison is again not good enough to constrain the best microscopic interactions and the corresponding EoS. It is also apparent that the calculations fail to reproduce the experimental data beyond 700 MeV/nucleon. This is an unexpected result because the experimental data seem to indicate that the nuclear transparency increases at large energies. {  In this case, the 5 smallest $\chi^2$ comparison with the data correspond to the SLY7, SKI4, BSR14, SKP and SKB interactions. They yield an average slope parameter $\left<L\right> = (43.6 \pm 20.0)$ MeV and an average neutron skin of $\left< \Delta r_{np}\right>=(0.0348\pm 0.0081)$ fm for the $^{28}$Si nucleus. There are currently no other experiments dedicated to study the neutron skin of $^{28}$Si than the ones based on reaction cross sections and charge-changing cross sections. It is clear from the experimental data accumulated so far that more experimental efforts are needed.}

\begin{figure}[t]
\begin{center}
\includegraphics[width=3.2in]{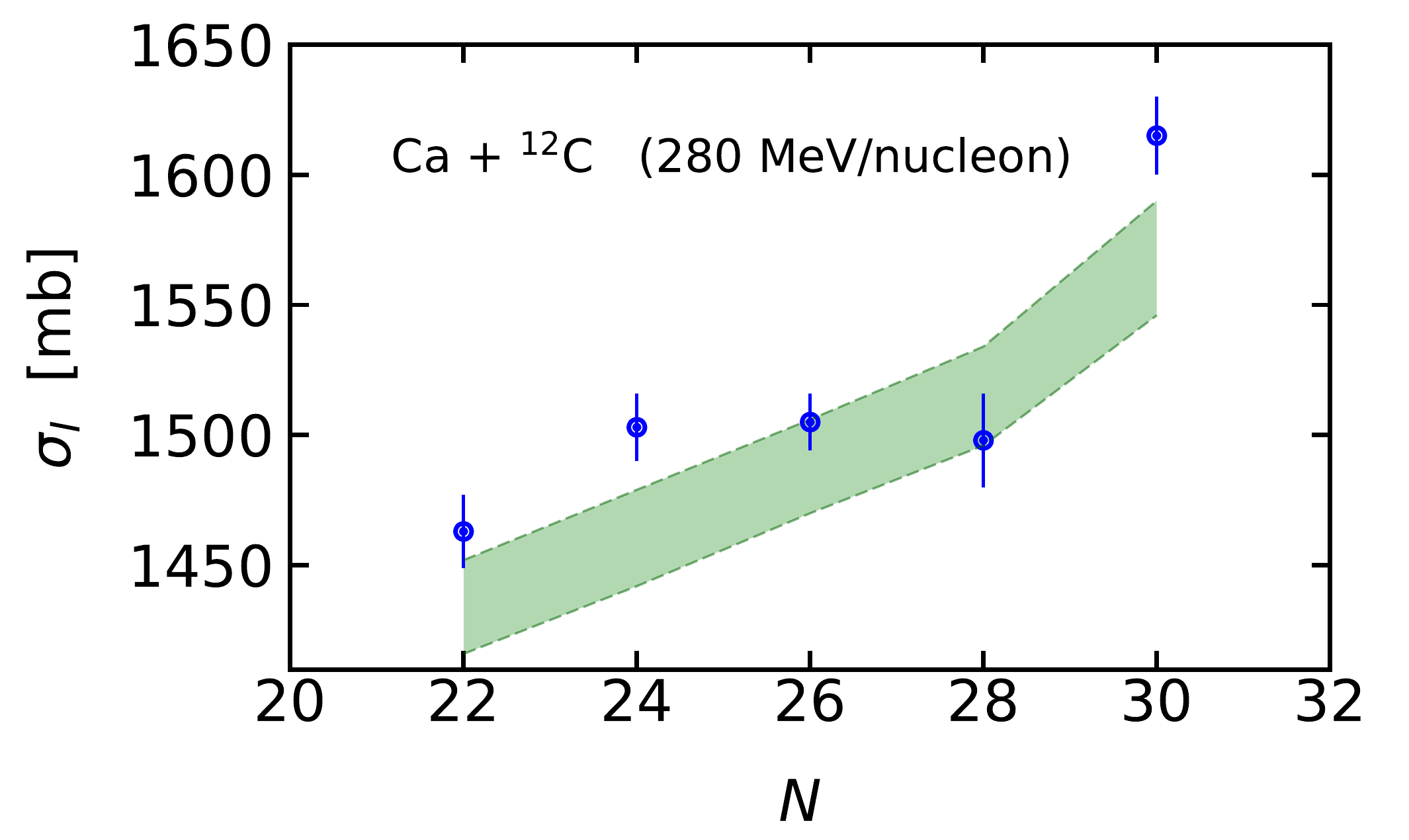}
\includegraphics[width=2.9in]{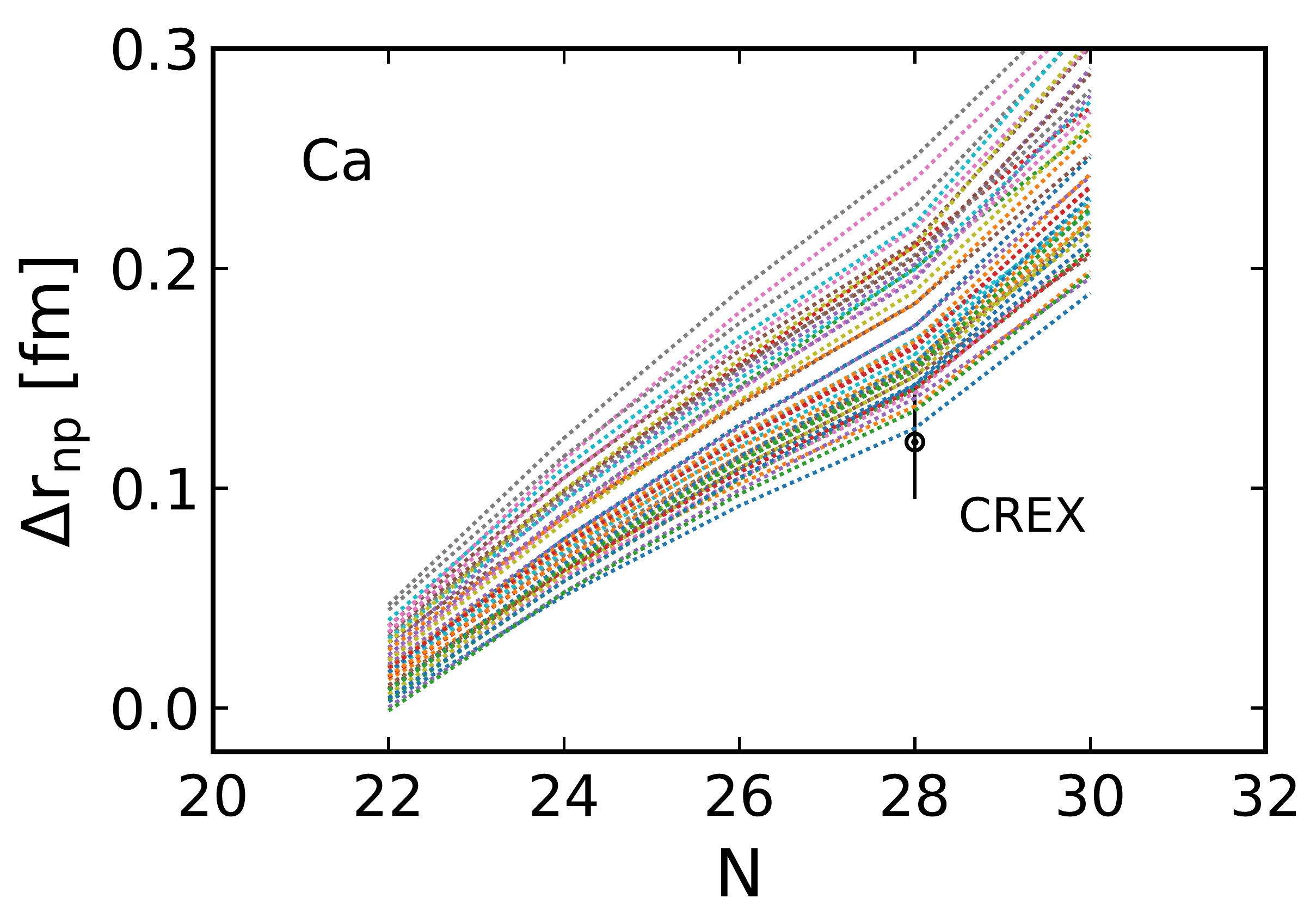}
\caption{User panel: Calculated interaction cross section $\sigma_{I}$ at 280 MeV/nucleon for the reaction Ca + $^{12}$C and different calcium isotopes compared to experimental data from Ref. \cite{PhysRevLett.124.102501}. The shaded band represents  the spread of calculations using {\color{black} calcium} densities obtained with non-relativistic and relativistic mean-field calculations with a total of 31 different interactions.
Lower panel: The neutron skin for calcium isotopes  calculated with the different mean-field models. The data from the CREX experiment for $^{48}$Ca \cite{CREXJlab} is also shown. }
\label{Canr}
\end{center}
\end{figure}

\begin{figure}[t]
\begin{center}
\includegraphics[
width=3.2in
]
{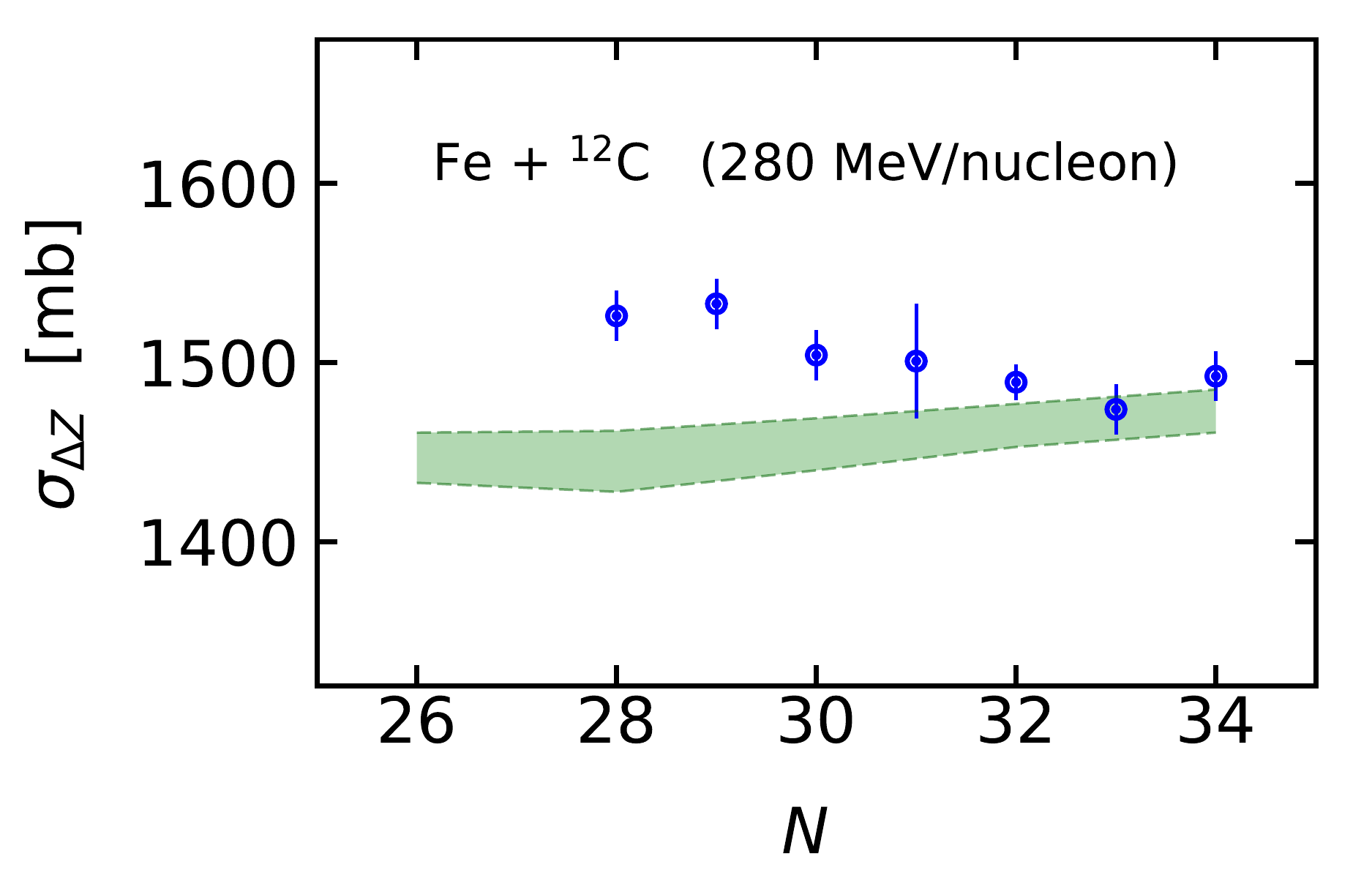}
\caption{Calculated charge changing cross section $\sigma_{\Delta Z}$ at 280 MeV/nucleon for the reaction Fe + $^{12}$C  and different iron isotopes compared to experimental data from Ref. \cite{YAMAKI2013774}. The shaded band represents the spread of calculations using  densities obtained with non-relativistic and relativistic mean-field calculations with a total of 31 different interactions.}
\label{Fenr}
\end{center}
\end{figure}

In Figure \ref{Canr}, upper panel, we shown the calculated interaction cross section, $\sigma_{I}$, for the reaction Ca + $^{12}$C and different calcium isotopes compared to experimental data from Ref. \cite{PhysRevLett.124.102501}. The shaded band represents  the spread of calculations using densities obtained with non-relativistic and relativistic mean-field calculations with a total of 31 different interactions. The lower panel displays the neutron skin for calcium isotopes  calculated with the different mean-field models. The data from the CREX experiment for $^{48}$Ca \cite{CREXJlab} is also shown.In Figure \ref{Fenr} we show the calculated charge changing cross section $\sigma_{\Delta Z}$ at 280 MeV/nucleon for the reaction Fe + $^{12}$C  and different iron isotopes compared to experimental data from Ref. \cite{YAMAKI2013774}. The shaded band represents the spread of calculations using densities obtained with non-relativistic and relativistic mean-field calculations with a total of 31 different interactions. {\color{black} In both cases, it is clear that the measurements for interaction cross sections and for charge changing cross sections are still not accurate enough to extract accurate information on neutron skins by comparison with theoretical predictions for the cross sections. In particular, for the data presented in Figure  \ref{Fenr}, the agreement with theory is very poor. The reason for these differences are not well understood. In the case of Ca + $^{12}$C collisions presented in Figure  \ref{Canr}, one is tempted to extract numbers for the $^{48}$Ca nucleus, where a reasonable agreement with experimental data is found,  to compare with the findings of the CREX experiment \cite{CREXJlab}. The HFB9 interaction yields the closest value for the interaction cross section, predicting a neutron skin of $ \Delta r_{np}=0.156$ fm and a slope parameter $L=39.8$ MeV. The second best result is obtained with the SKX interaction, yielding  $ \Delta r_{np}=0.167$ fm and  $L=33.2$ MeV. These results for the neutron skin are about 20\% higher than the value $\Delta r_{np}=0.121 \pm 0.026$ fm reported by the CREX experiment  \cite{CREXJlab} and, in contrast, the slope parameter $L$ derived from these two interactions are also much smaller than the value $L\simeq 119$ MeV obtained with the NL3 interaction \cite{PhysRevC.55.540} and favored by the CREX experiment \cite{CREXJlab}.}

\subsection{Comparison to past and presently proposed methods}

It is clear from our limited analysis that the present status of the experimental data on fragmentation reactions is not yet at the level of accuracy to allow a good constraint on the best possible microscopic interactions and in the process to obtain the best possible EoS for neutron stars. Other experimental methods have been devised in the past but also lack the necessary accuracy needed to constrain the best possible EoS. 

Recently, a new method was introduced, with the neutron skin extracted from an analysis of the parity violation effect on electron scattering off lead and calcium nuclei \cite{DONNELLY1989589,PhysRevC.63.025501,Thiel-2019}. This method would in principle allow a measurement  independent of the complications of the strong force.  As we discussed before, a recent experiment based on this technique were performed at the Jefferson Laboratory \cite{PhysRevLett.126.172502}  with a $^{208}$Pb target and found a large neutron skin  of $\Delta r_{np} = 0.283 \pm 0.071$ fm, implying a slope parameter of $L=106 \pm 37$ MeV, larger than expected from most microscopic calculations and also from other experiments with the same nucleus. For example, Ref. \cite{ISI:000293447900005} reported a value of $\Delta r_{np} = 0.156^{ + 0.025}_ {- 0.021}$ fm. Moreover, astronomical observations from  NICER and LIGO/Virgo collaborations are compatible with much smaller values for  $\Delta r_{np}$ and $L$. Although the JLab experiment  was successful, statistical significance was not achieved \cite{PhysRevLett.126.172502}. It has been advertised by the JLab \cite{CREXJlab} that the CREX experiment, a variant of the PREX experiment but this time using $^{48}$Ca as target, has obtained a much smaller value of the neutron skin, $\Delta r_{np}=0.121 \pm 0.026$ fm. Hence, the PREX and the CREX experiments seem to provide incompatible results for the size of the neutron skin.

\begin{figure}[t]
\begin{center}
\includegraphics[
width=3.5in
]
{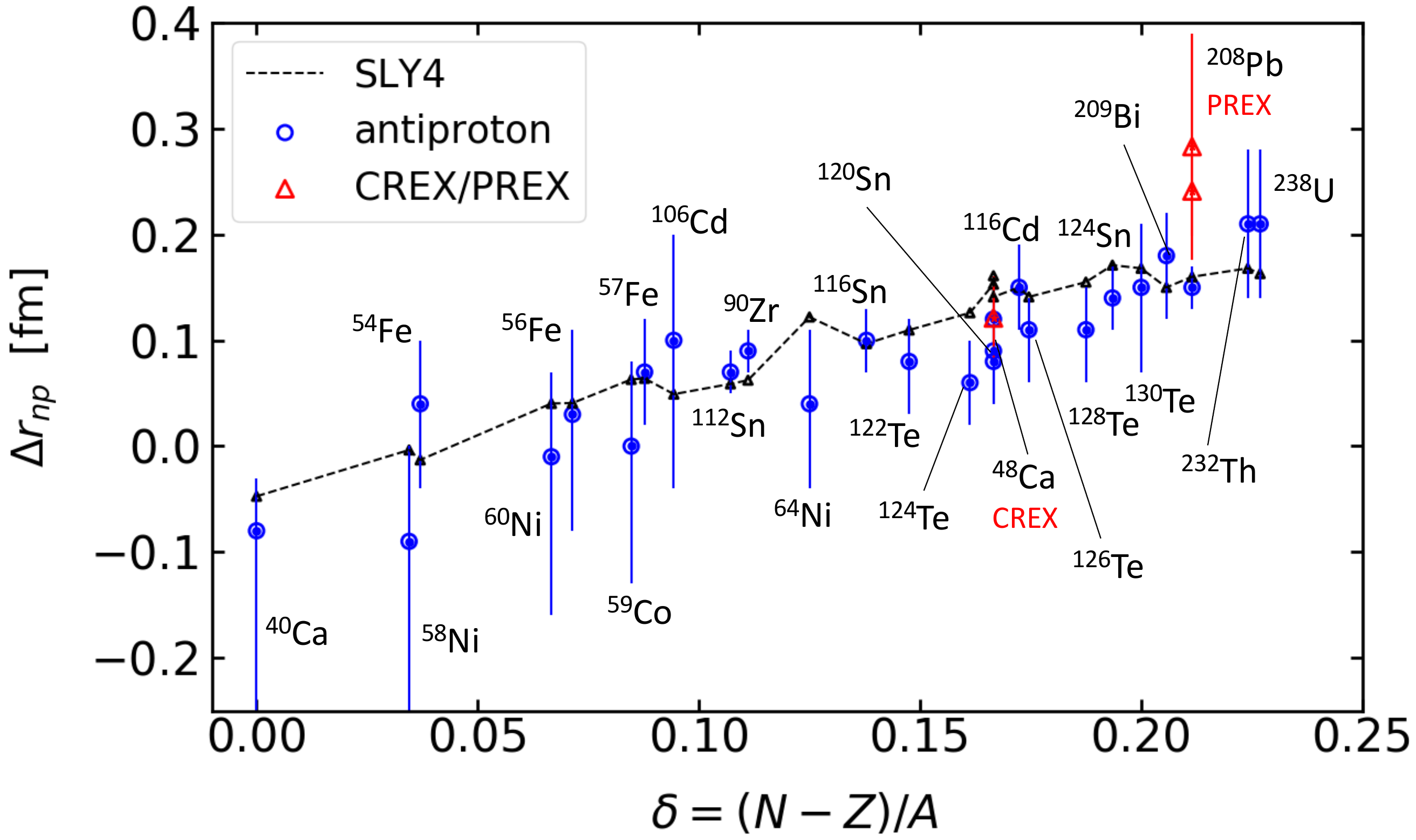}
\caption{{  Neutron skin thickness as a function of the   isospin asymmetry parameter $\delta=(N-Z)/A$ compared to  experimental data extracted from antiproton annihilation data  \cite{PhysRevLett.87.082501,doi:10.1142/S0218301304002168} and with data from the CREX experiment for $^{48}$Ca \cite{CREXJlab} and for the  PREX experiment for $^{208}$Pb (two data points) \cite{PhysRevLett.126.172502,PhysRevLett.108.112502}. The  small triangles guided by a dashed line correspond to the prediction of the SLY4 interaction in the HFB formalism.} }
\label{drnpall}
\end{center}
\end{figure}

{  In Figure  \ref{drnpall} we show the neutron skin $\Delta r_{np}$ of several nuclei calculated with the SLY4 interaction as a function of the isospin asymmetry parameter $\delta = (N-Z)/A$.  The experimental values were extracted from antiproton annihilation data \cite{PhysRevLett.87.082501,doi:10.1142/S0218301304002168} and with data from the CREX experiment for $^{48}$Ca \cite{CREXJlab} from the PREX experiment for $^{208}$Pb (two data points)  \cite{PhysRevLett.126.172502,PhysRevLett.108.112502}.}  It is evident that, as expected, both the experimental data and theory display a trend of increasing neutron skin with increasing  isospin asymmetry parameter $\delta = (N-Z)/A$ for all nuclei studied so far. {  It is also clear from the plot that the popular SLY4 interaction, as many others, yield a pretty reasonable description of the average behavior of the neutron skin of most studied nuclei with the antiproton and electron PV scattering techniques. The PREXI  \cite{PhysRevLett.126.172502} and PREXII  \cite{PhysRevLett.108.112502} data are hard to explain and are in tension with the microscopic calculations. }  {  It is worthwhile mentioning that the extraction of neutron skins using antiprotons is more of a direct technique for this purpose. The antiproton-nucleon interaction is very strong and antiprotons interacting with nuclei are absorbed and annihilate already at the nuclear periphery, where the nucleon density is significantly smaller than the central nuclear density.  One uses radiochemical methods \cite{JASTRZEBSKI1993405}, supplemented with radiochemical data \cite{PhysRevC.58.3195} by means of in-beam antiprotonic x-ray studies \cite{PhysRevLett.87.082501,doi:10.1142/S0218301304002168} thus determining strong-interaction level widths and shifts in isotopically enriched targets. The major hurdle in analyzing the antiproton data is the theoretical description of antiproton-nucleus interactions (optical potential) \cite{FRIEDMAN2005283}.  A renewed interest in theoretical derivations of the antiproton-nucleus optical potential has recently emerged to support future experiments \cite{Lazau2022}. In contrast, both electron PV scattering experiments and nuclear fragmentation data analysis are based on a comparison with theory input for matter density obtained with microscopic calculations either through neutron density form factors, or Glauber nucleon knockout profile functions. Therefore, the three techniques described here are very different, complementary, but strongly dependent on theory input. It is also clear that one needs to increase the number, accuracy and methods used in the experiments to narrow down the proper EoS of nuclear matter. }

\begin{figure}[t]
\begin{center}
\includegraphics[
width=3.5in
]
{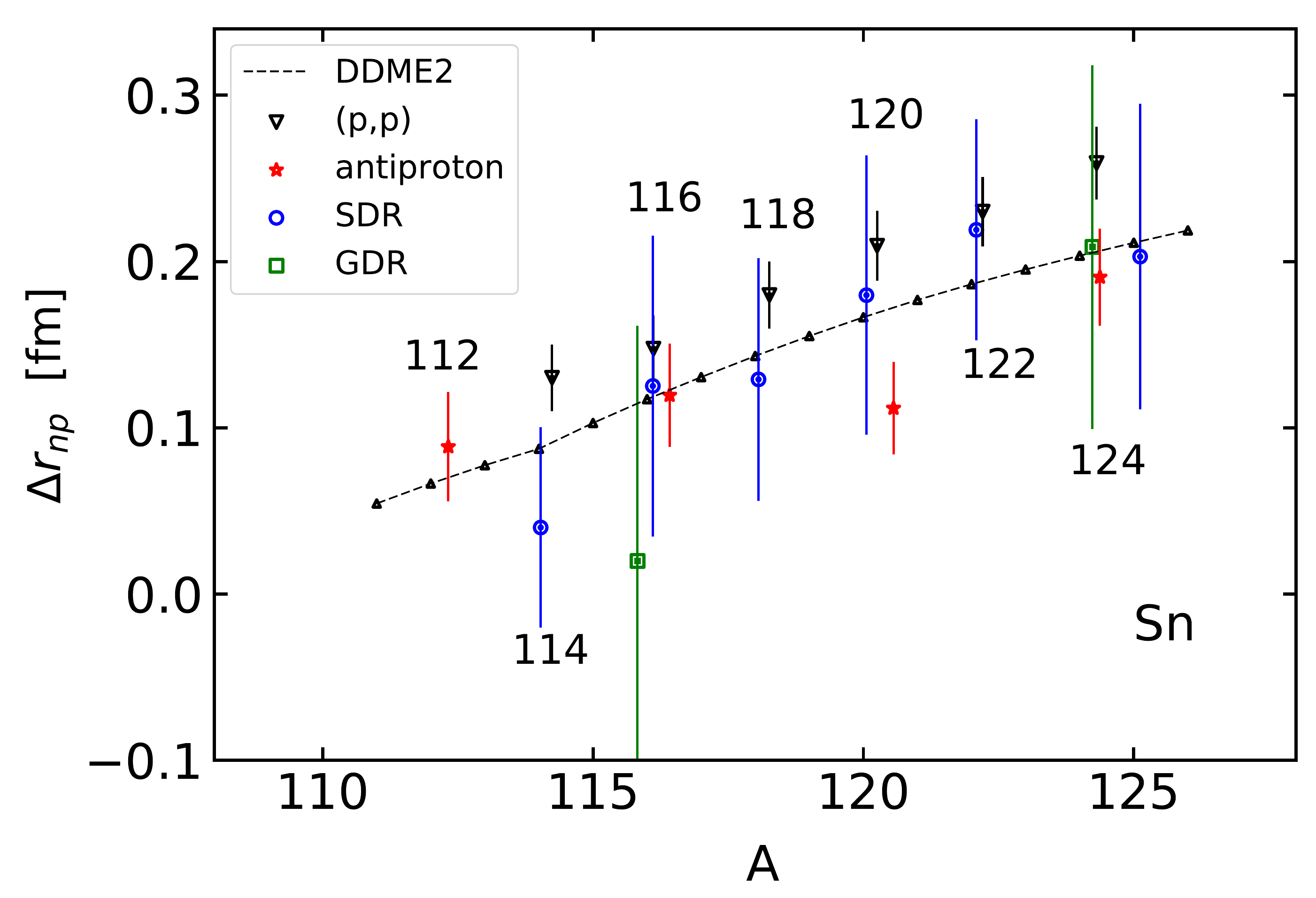}
\caption{{  Neutron skin $\Delta r_{np}$ of Sn isotopes calculated with the DDME2 interaction in the RMF formalism. The experimental data Sn were measured with (p, p) reactions (triangles)
\cite{PhysRevC.19.1855}, antiproton atoms (stars) \cite{PhysRevLett.87.082501}, giant dipole resonance method (squares) \cite{CSATLOS2003C304} and spin dipole resonance method (circles) \cite{PhysRevLett.82.3216,KRASZNAHORKAY2004224} To avoid superposition of the data, we displace  them slightly and label each group according to their masses $A$.}  
}
\label{drnpSn}
\end{center}
\end{figure}

A promising method to extract the nuclear EoS is still based on the effect on the nucleus-nucleus collisions of the growth of the neutron skin along an isotopic chain. This method can be tested in different experiments using hadronic probes and a plethora of techniques such as elastic and inelastic scattering, photo absorption, fragmentation, etc. In particular, radioactive beam facilities can be very helpful, as they provide isotopic beams with different neutron/proton ratios.  In Figure  \ref{drnpSn} we present the neutron skin $\Delta r_{np}$ of Sn isotopes calculated with the DDME2 microscopic interaction. {  The experimental data for Sn nuclei were measured with (p, p) reactions (triangles)
\cite{PhysRevC.19.1855}, antiproton atoms (stars) \cite{PhysRevLett.87.082501}, giant dipole resonance method (squares) \cite{CSATLOS2003C304} and spin dipole resonance method (circles) \cite{PhysRevLett.82.3216,KRASZNAHORKAY2004224}. The small triangles and dashed line displays the numerical calculations for the neutron skin, $\Delta r_{np}$, calculated with the DDME2 relativistic mean field interaction. This interaction predicts a slope parameter $L=51$ MeV.}  It is noticeable that the distinct probes of the neutron skin seem to yield inconsistent results, requiring a further  assessment of the experiments systematics. The small error bars and apparent well behaved trend of the experimental data based on polarized proton scattering \cite{PhysRevLett.47.1436,PhysRevLett.87.082501} seem to provide a promising way to move forward with radioactive beams and indirect techniques. {  The same sort of comparison of more accurate data  along an isotopic chain  with theory prediction as in Fig.  \ref{drnpSn} will provide another useful constraint on the symmetry energy content of the ANM EOS.}

The experiments using fragmentation reactions can and will certainly help in determining which additional information that can be gleaned from the experiments. In particular, fragmentation reactions can be performed using combinations of numerous projectiles and targets \cite{AumannPRL119.262501}. Using  proton and carbon targets have the advantage of testing the skin with a deeply penetrating probe (proton) and a more surface oriented one (carbon). Light targets are preferable to avoid large contributions from fragmentation due to Coulomb excitation. Fragmentation reactions can also use projectiles with a varying combination of neutron to proton ratio, in particular with radioactive beam facilities. Tin, for example, has a long isotopic chain and is one of the candidates for testing an isospin dependence of  the neutron skin $\Delta r_{np}$. In Table \ref{table2} we  summarize the calculations for the total neutron removal cross sections using $^{12}$C and proton targets for a selected number of tin isotopes and energies compatible with the GSI  facility in Germany.   {  Here we use the SLY4 interaction in the HFB formalism. For this interaction, the slope parameter is $L=45.9$ MeV the neutron skins and are: (a) $^{124}$Sn, $ \Delta r_{np} = 0.171$ fm, (b) $^{128}$Sn, $ \Delta r_{np} = 0.198$ fm, (c)  $^{132}$Sn, $ \Delta r_{np} = 0.222$ fm, and (d)  $^{134}$Sn, $ \Delta r_{np} = 0.250$ fm. The increase of the skin as a function of the neutron-proton asymmetry attests the value of studying the neutron skin evolution  along an isotopic chain to better constrain the best interaction to fit the data.}

\begin{center}
\begin{table}[h]
\caption{Total neutron removal cross sections for Sn projectiles incident on carbon and hydrogen targets {  using the SLY4 interaction in the HFB formalism}. The corrections due to Pauli blocking,  Coulomb repulsion (Coul), and excitation of collective (GDR+GQR) states are added in sequence.}
\setlength{\tabcolsep}{1.16mm}
{
\begin{tabular}{lccccccc} \hline \hline
          & $E_{beam}$  & Target  &$\sigma_{\Delta N}$& (+ Pauli)&  (+ Coul) & (+ GR) \\  \hline
    & MeV/nucl & &mb& mb&mb&mb\\  \hline
    $^{124}$Sn & 400   &C & 419&429  &  427 &493 \\
    &    &H & 254&263 &  263 & 271 \\
     & 900   &C & 420&425 &  424 &469  \\
    &    &H & 268&273 &  272 & 279 \\
    $^{128}$Sn & 900   &C & 449&455 &  454 & 516 \\
    &    &H & 288&291 &  290  & 297\\    
   $^{132}$Sn      & 600   &C & 472&481   &  480 &  541 \\   
    &    &H & 365&370 &  367 & 374 \\    
     & 900   &C & 476&481 &  480 & 542 \\
         &    &H & 306&309 &  308 & 315 \\         
   $^{134}$Sn      & 600   &C & 496&505 &  504 & 563  \\ 
    &    &H & 378&383 &  381 & 387  \\  
     & 900   &C & 501&506 &  505 & 565  \\     
    &    &H & 321&323 & 322 & 328  \\
\hline \hline
\end{tabular}}\label{table2}
\end{table}
\end{center}

From Table \ref{table2} one sees that the most challenging correction to the rather clean and commonly used theoretical picture presented by the Glauber-like method described by Eqs. (\ref{abrasionp2}-\ref{sigmaDN}) is the excitation of collective modes leading to neutron emission. This has been rarely discussed in the literature in this context, except in a few publications \cite{Bertulani.PRC.100.015802}. But using both proton and carbons targets will help to disentangle the effects of this correction. Performing experiments at different energies will also help to constrain the dependence of the cross sections on the neutron and proton nuclear densities. {  On the other hand, in inclusive fragmentation experiments as proposed in Ref.  \cite{R3Bneutronremoval}, the total neutron removal cross section could be determined with a precision of around 1\%. In order to avoid the theoretical uncertainty of calculation the contribution due to collective excitations as mentioned above, a recently proposed experiment will directly determine in addition this part of the cross section, which contributes to about 10\% to the total cross section. If neutron removal is caused only by neutron evaporation after excitation, the neutrons will be detectable at small angles in the laboratory around the beam axis (due to the high beam velocity). The total neutron-changing cross section minus this part can be directly compared to Glauber calculations using different theoretical densities (corresponding to different $L$ values and neutron skins). By this comparison, the $L$ value and $\left< r^2\right>_n$ will be constrained. A detailed sensitivity study for this approach is discussed in Ref. \cite{AumannPRL119.262501}. The uncertainty of the extracted value will contain besides the experimental error the uncertainty of the Glauber model used.}

\section{Conclusions}

In this work we have made an analysis of the contributions of numerous physics effects on the interaction, reaction, charge-changing and neutron-changing cross sections. We have studied the contribution of (a) Pauli-blocking, (b) Fermi motion, (c) eikonal corrections, (d) Coulomb recoil, (d) Coulomb and (e) nuclear excitation of giant resonances. Clear discrepancies with published experimental data have been found for interaction, reaction, charge-changing and neutron-changing cross sections. Notably, the data reported in the literature for charge-changing reactions seem to favor a transparency scenario for the cross sections at the largest energies. This is not easy to understand, as the nucleon-nucleon cross sections are rather energy independent at these energies.

It is also clear that the data for total neutron removal, or neutron-changing, cross sections are not yet accurate enough to provide a tight constraint on the mean-field calculations and the corresponding EoS.  In particular, we have observed in our calculations that old interactions such as the SKIII, SKA and SKB interactions \cite{BEINER197529,KOHLER1976301} fare better in reproducing some of the published data than more modern and celebrated interactions such as the SLY, DDME2, or UNEDF interactions \cite{CHABANAT1998231,STOITSOV20131592,PRC71-024312}.  {  No single interaction stands out as a better one to reproduce the experimental data that we have discussed in this work}. At this stage, a global fit or machine learning approach to constrain the parameters of the best theory does not seem to be very useful in view of the scarcity and inconsistency of the data reported in the literature. Also, the celebrated PREX experiment has not been conclusive, yielding an unexpectedly large neutron skin for lead. {  The CREX experiment has obtained a much smaller neutron skin than the expected by an extrapolation of the PREX experiment.}  

The recently proposed experimental campaign to determine neutron-changing cross sections in inverse kinematics for projectiles with a large range of isospin asymmetry  \cite{AumannPRL119.262501} will help to increase the available data to assess the dependence of the neutron skin on the isospin asymmetry parameter $\delta=(N-Z)/A$. These experiments will have a direct impact on the determination of the symmetry term of the EoS.

\medskip

{\bf Acknowledgments.} E.A.T. acknowledges  partial support by the Coordena\c c\~ao de Aperfei\c coamento de Pessoal de N\' \i vel Superior - Brazil (CAPES). T.A. acknowledges support by the German Federal Ministry of Education and Research (BMBF, project 05P2015RDFN1). C.A.B. acknowledges support by the U.S. DOE grant DE-FG02-08ER41533 and the Helmholtz Research Academy Hesse for FAIR. B.V.C. acknowledges support from grant 2017/\-05660-0 of the S\~ao Paulo Research Foundation (FAPESP), grant 303131/2021-7 of the CNPq and the INCT-FNA project 464898/2014-5.


\end{document}